\documentclass[aps,twocolumn,nofootinbib, superscriptaddress, floatfix]{revtex4-1}

\usepackage{placeins}
\usepackage{soul}
\usepackage[usenames,dvipsnames,svgnames]{xcolor}
\definecolor{redd}{rgb}{0.8, 0.1,0.2}
\definecolor{navy}{rgb}{0.05, 0.23,0.75}
\usepackage[colorlinks]{hyperref}
\usepackage{chngcntr}
\hypersetup{
     colorlinks   = true,
     citecolor    = navy,
	linkcolor = redd,
	urlcolor=navy,
	anchorcolor=blue
}
\usepackage{bbm}
\usepackage{amsfonts}
\usepackage{amsmath}
\usepackage{mathrsfs}
\usepackage{soul}

\usepackage{amsmath,amssymb,amsbsy,bm}

\usepackage{epsfig}
\usepackage{graphicx}
\usepackage{wrapfig}                 
\usepackage{url}
\usepackage{hyperref}
\usepackage{float}
\usepackage{pstricks}
\usepackage{color}
\usepackage{multirow}
\usepackage{lipsum}
\usepackage{enumitem}
\usepackage{ctable}
\newcolumntype{L}{>{\centering\arraybackslash}m{1.5cm}}

\usepackage{titlesec}
\titleformat{\section}[runin]{\normalfont \bfseries}{\thesection}{1em}{}
\titleformat{\subsection}[runin]{\normalfont \bfseries}{\thesubsection}{1em}{}

\usepackage{enumitem}

\usepackage{chngcntr}

\newcommand{\be}{\begin{equation}}
\newcommand{\ee}{\end{equation}}
\newcommand{\bea}{\begin{eqnarray}}
\newcommand{\eea}{\end{eqnarray}}
\newcommand{\beq}{\begin{equation}}
\newcommand{\eeq}{\end{equation}}

\newcommand{\twiddle}{{\raise.17ex\hbox{$\scriptstyle\sim$}}}

\begin{document}

\title{Higgscitement: Cosmological Dynamics of Fine Tuning}
\author{Mustafa A. Amin} 
\affiliation{Physics \& Astronomy Department, Rice University, Houston, Texas 77005, USA}
\author{JiJi Fan} 
\affiliation{Department of Physics, Brown University, Providence, RI, 02912, USA}
\author{Kaloian D. Lozanov} 
\affiliation{Max Planck Institute for Astrophysics, Karl-Schwarzschild-Str. 1, 85748 Garching, Germany}
\author{Matthew Reece}
\affiliation{Department of Physics, Harvard University, Cambridge, MA, 02138, USA}

\begin{abstract}

The Higgs potential appears to be fine-tuned, hence very sensitive to values of other scalar fields that couple to the Higgs. We show that this feature can lead to a new epoch in the early universe featuring violent dynamics coupling the Higgs to a scalar modulus. The oscillating modulus drives tachyonic Higgs particle production. We find a simple parametric understanding of when this process can lead to rapid modulus fragmentation, resulting in gravitational wave production. A nontrivial equation-of-state arising from the nonlinear dynamics also affects the time elapsed from inflation to the CMB, influencing fits of inflationary models. Supersymmetric theories automatically contain useful ingredients for this picture.

\end{abstract}

\maketitle

%\section{Introduction}
%%%%%%%%%%%%%%%%%%%%%%%%%%%%%%%%%%%%%%%%

\section{Introduction}

The origin of the Higgs mass and the mechanism of electroweak symmetry breaking (EWSB) are among the biggest puzzles in fundamental physics. The Higgs mass receives large quantum corrections unless there is new physics to tame this sensitivity, e.g.~supersymmetry (SUSY), which predicts that the masses of the Higgs and numerous other scalars lie near a common scale of supersymmetry breaking. The Large Hadron Collider (LHC) has not found the predicted plethora of new particles near the Higgs mass. However, this does not rule out a scenario such as SUSY. Our universe may simply lie in the region of parameter space where the Higgs boson is {\it accidentally} much lighter than the other scalars. In this article, we will show that such a scenario can lead to dramatic non-perturbative dynamics in the early universe, generating potentially observable cosmological signatures. 

Within SUSY extensions of the SM, the parameters of the SM are not truly constant but are affected by the values of scalar fields called {\em moduli}. These fields have couplings to the SM suppressed by a large scale (e.g., the Planck scale), so they cannot be produced or detected at colliders. Our vacuum is a minimum of the potential for the moduli and the Higgs. In this context, the LHC results hint that this minimum lies near a critical point in parameter space where the Higgs is significantly lighter than the typical SUSY scale, with the Higgs potential precariously balanced between no EWSB and severe EWSB. Can we test this scenario with cosmology?

In the present universe, we cannot vary parameters to explore the potential near such a critical point. However, the early universe might already have carried out such an exploration. In the early universe, moduli were generically displaced from the minimum and evolving in time. We will show that the accidental lightness of the Higgs in the present universe or equivalently, the weakness of EWSB, can potentially lead to non-perturbative, violent cosmological dynamics of the moduli and SM fields. Such dynamics can yield potentially observable signatures like a high frequency $\sim 10\,\rm kHz$ stochastic gravitational background and change the expansion history of the universe. 

Testing whether we live in a ``meso-tuned" universe is a key goal for a future very high energy hadron collider \cite{Arkani-Hamed:2015vfh}.  Our goal here is to explore the cosmological dynamics, seeking signals that give a positive and direct test of fine tuning, separate from collider probes. We intend to open a new angle on the possible connection between EWSB and early universe cosmology.\footnote{Related work includes studies of: time-dependent SM parameters in the early universe \cite{Fardon:2003eh, Servant:2014bla, Graham:2015cka, Ema:2016ehh,  Baldes:2016rqn, vonHarling:2016vhf}; a different possible inflationary probe of fine-tuning \cite{Kumar:2017ecc}; whether parametric resonance can solve the moduli problem \cite{Shuhmaher:2005mf} or not \cite{Giblin:2017wlo}.}

Our goal here is not to {\em explain} fine tuning. Plausibly there is a landscape of possible theories with varying amounts of fine tuning, and we find ourselves in a moderately tuned vacuum. (A separate tuning is needed to cancel the cosmological constant.) For our purposes it does not matter whether this is due to random chance or anthropic selection; what matters for us are the cosmological implications of this tuning. Our work explicitly demonstrates that instead of being merely an aesthetic concept, fine-tuning has concrete physical consequences. 

%~~~~~~~~~~~~~~~~~~~
\begin{figure}[b!] %  figure placement: here, top, bottom, or page
   \centering
   \includegraphics[width=3.2in]{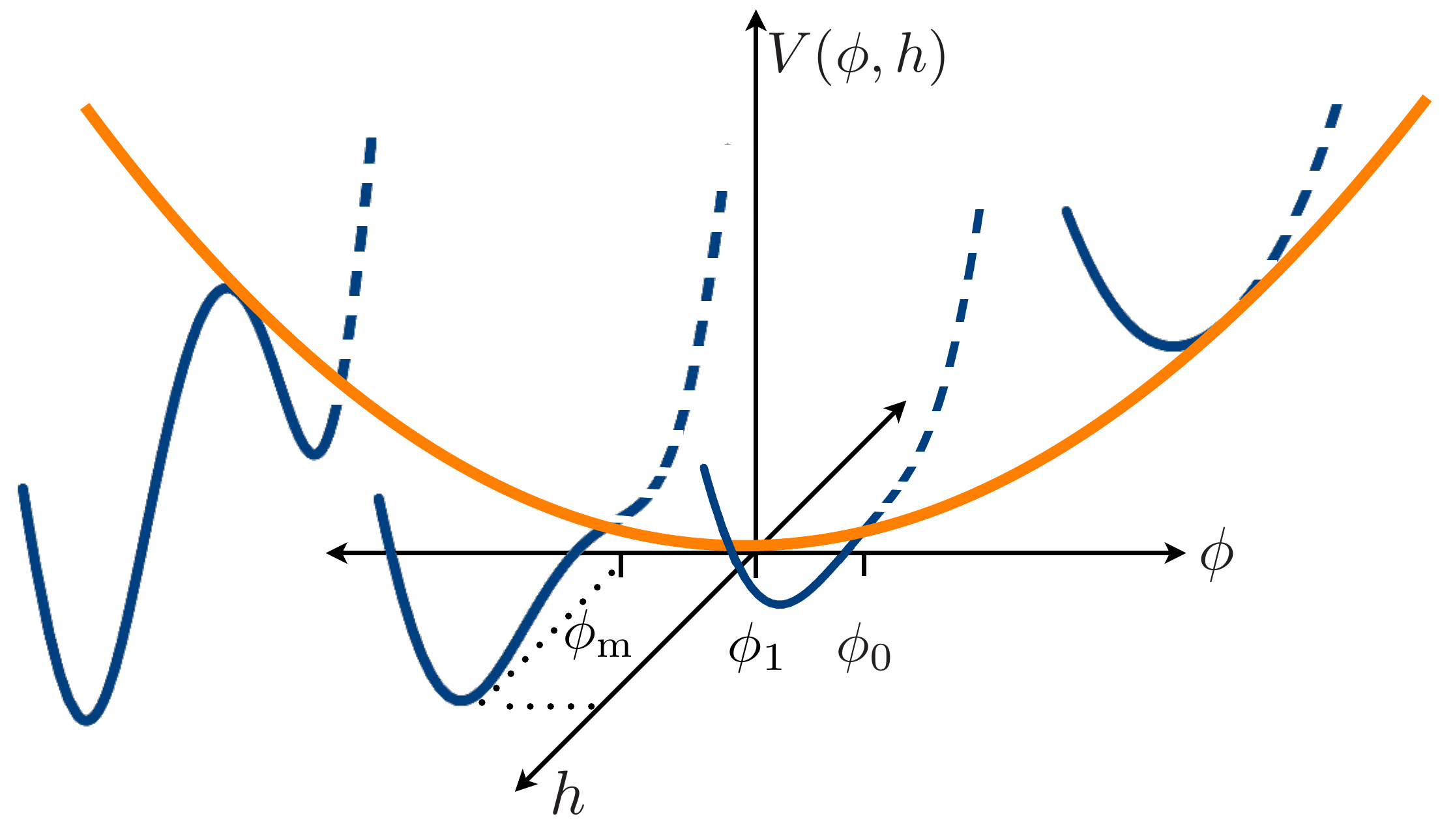} 
   \caption{The shape of the Higgs-moduli potential. The global minimum of the potential is at $(\phi=\phi_{\rm m},h\ne0)$, whereas $\phi_0$ is the point of symmetry breaking. %The minimum of the potential for $h=0$ is at $\phi=\phi_1$.
   }
   \label{fig:Pot}
\end{figure}
%~~~~~~~~~~~~~~~~~~~

%%%%%%%%%%%%%%%----- A Simple Model -------%%%%%%%%%
\section{A Simple Model}  \label{sec:simplemodel}
%%%%%%%%%%%%%%%%%%%%%%%%%%%%%%%%%%%%%

We seek a simplified model capturing the assumption that a Higgs field $h$ couples to a modulus $\phi$ (with characteristic field range $f$) such that for typical values of $\phi$, the Higgs mass takes a natural value of order $M$ (e.g.~the SUSY-breaking scale), but for particular values of $\phi$ the Higgs mass may be much smaller. Such a potential could have the form
\begin{align}
\frac{1}{2} m_\phi^2 (\phi - \phi_1)^2 + M^2 \frac{\phi - \phi_0}{f} h^\dagger h + \lambda(h^\dagger h)^2 + V_0.
\label{eq:toymodel} 
\end{align}
A priori we expect $\phi_0 \sim \phi_1 \sim f$. The value $\phi = \phi_0$ is the point of marginal EWSB, whereas $\phi=\phi_1$ is the point where $V$ is minimized for $h=0$ (along the ``ridge" in the potential in Fig.~\ref{fig:Pot}). The global minimum of this potential is at $\phi_{\rm m}=(b\phi_0-\phi_1)/{(b-1)}$ and  $|h_{\rm m}|= M\sqrt{(\phi_0-\phi_1)/(2\lambda f (1-b))}$, where 
\be
b\equiv \frac{M^4}{2\lambda f^2 m_\phi^2}<1\,.
\label{eq:bDef}
\ee
The parameter $b$ plays a critical role in the dynamics; $b < 1$ is necessary for the potential to be bounded from below.

There is no a priori reason why the the global minimum of the potential lies near the point of marginal EWSB. The closer $\phi_{\rm m}$ is to $\phi_0$, the greater our surprise. This means,
\be
{\rm Fine~tuning} \Leftrightarrow \Delta \equiv \frac{\phi_0-\phi_{\rm m}}{f}\ll 1\,.
\label{eq:finetuning}
\ee
In terms of this fine tuning parameter, the observed Higgs mass around the global minimum is 
\be
m_h^2=2M^2\Delta.
\ee
This is closely related to fine tuning in the usual sense: if eq.~\eqref{eq:toymodel} represented a {\em tree-level} potential, loop corrections, including a $\phi$ tadpole, would shift the minimum away from marginal EWSB and spoil the coincidence. However, we take eq.~\eqref{eq:toymodel} to represent the quantum-corrected effective potential, so that we do not have to compute loop-induced shifts in VEVs.

We will mostly have in mind fine-tuned supersymmetric theories, where this toy simplified potential can arise with $M^2 \sim m_{\rm soft}^2$ as explained in \S\ref{supp:modulicouplings}. For concreteness, we will focus on $\Delta=10^{-6}$ which corresponds to $M\sim 10^2\, \rm{TeV}$. We consider the hierarchy $m_{h}^2 \ll m_\phi^2 \lesssim M^2 \ll f^2$. Self-interaction terms which we have neglected for simplicity, e.g.~$(m_\phi^2 / f^2) \phi^4$ or $\frac{1}{f^2} \phi^2 \partial_\mu \phi \partial^\mu \phi$, could have important effects on the dynamics (such as oscillon formation \cite{Gleiser:1993pt,Copeland:1995fq,Amin:2011hj,Amin:2013ika,Antusch:2017flz}).  

For the aforementioned hierarchy of scales, the nonlinear dynamics we are interested in requires that $\lambda$ be much smaller than the SM value ($\lambda \sim 0.1$), but does not necessarily imply an inconsistency with the the observed electroweak properties (see \S\ref{sec:moremodel}). Alternatively, we will argue that the relevant nonlinear dynamics is still present with a different hierarchy and $\lambda\sim 0.1$, although it becomes extremely challenging to simulate numerically. We also note that for simplicity, our simulations substitute a real scalar field for the complex $h$.

%~~~~~~~~~~~~~~~~~~~
\begin{figure}[t!] %  figure placement: here, top, bottom, or page
   \centering
   \includegraphics[width=3.2in]{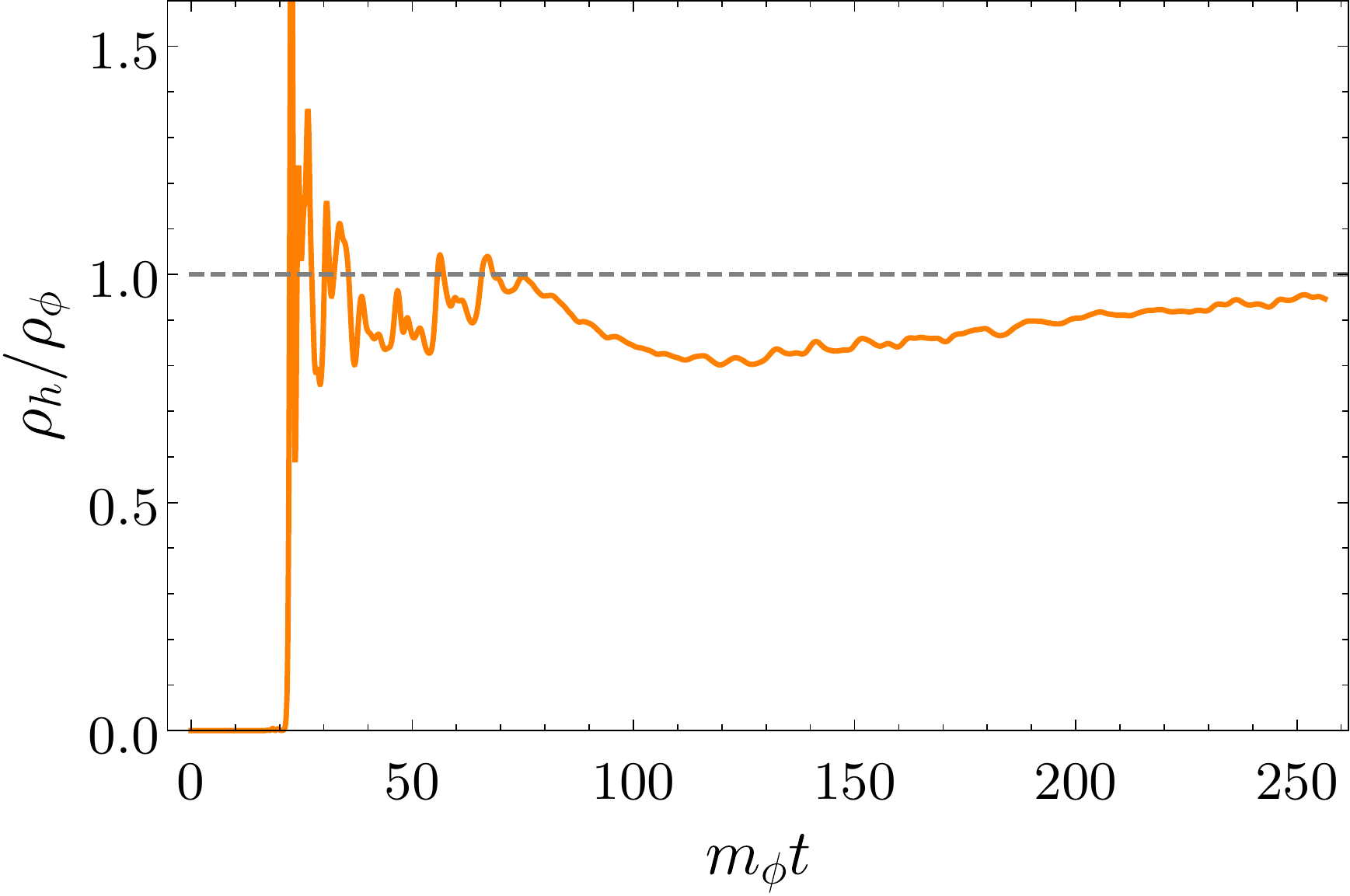} 
   \caption{The ratio of the spatially averaged energy density in the Higgs and modulus fields as a function of time, from our lattice simulations. This dynamics is representative of the energy transfer between the modulus and Higgs fields when $b\equiv M^4/2\lambda f^2 m_\phi^2\sim\mathcal{O}[1]$. For this plot we have chosen $\Delta =10^{-6}$, $M^2/m_\phi^2=10^2$, $M/f=10^{-13}$ and $\lambda\sim10^{-24}$,  which corresponds to $b=0.9$. We have confirmed that changing the parameters (for example, increasing $\lambda$ by $6$ orders of magnitude) while keeping $b\sim 1$ fixed, does not qualitatively change our results. 
   }
   \label{fig:b1r}
\end{figure}
%~~~~~~~~~~~~~~~~~~~

%~~~~~~~~~~~~~~~~~
\begin{figure*}[t] %  figure placement: here, top, bottom, or page
   \centering
      \includegraphics[width=6.6in]{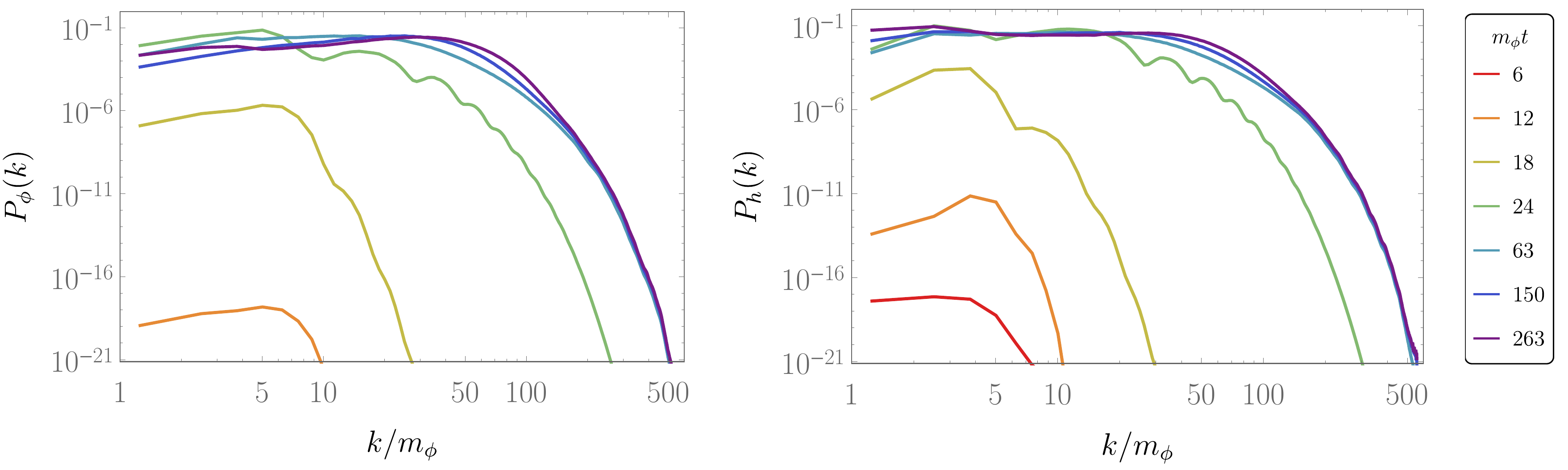} 
   \caption{The evolution of the normalized fields power spectra for the model with $\Delta =10^{-6},b =0.9,q=10^2, f=m_{\rm pl}$. The normalized power spectrum of a field $F(\bf{x})$ is $P_{F}(k)\equiv\phi_{\rm osc}^{-2}(d/d\ln k)\overline{F^2(\bf{x})}$, where $\phi_{\rm osc}$ is the amplitude of the background modulus oscillations. For this normalization, when $P_{\phi}(k)=\mathcal{O}(1)$, the modulus becomes inhomogeneous. Initially, the tachyonic instability in the Higgs is closely followed by excitations in the modulus (due to re-scattering). Comoving modes $k<m_{\phi}q^{1/2}$ grow exponentially. At the third oscillation of the modulus backreaction takes place. The spectra then settle down and power slowly propagates towards higher comoving modes.}
   \label{fig:ps}   
\end{figure*}
%~~~~~~~~~~~~~~~~~

%%~~~~~~~~~~~~~~~~~~~
%
\begin{figure*}[t] %  figure placement: here, top, bottom, or page
\centering
\includegraphics[width=7.1in]{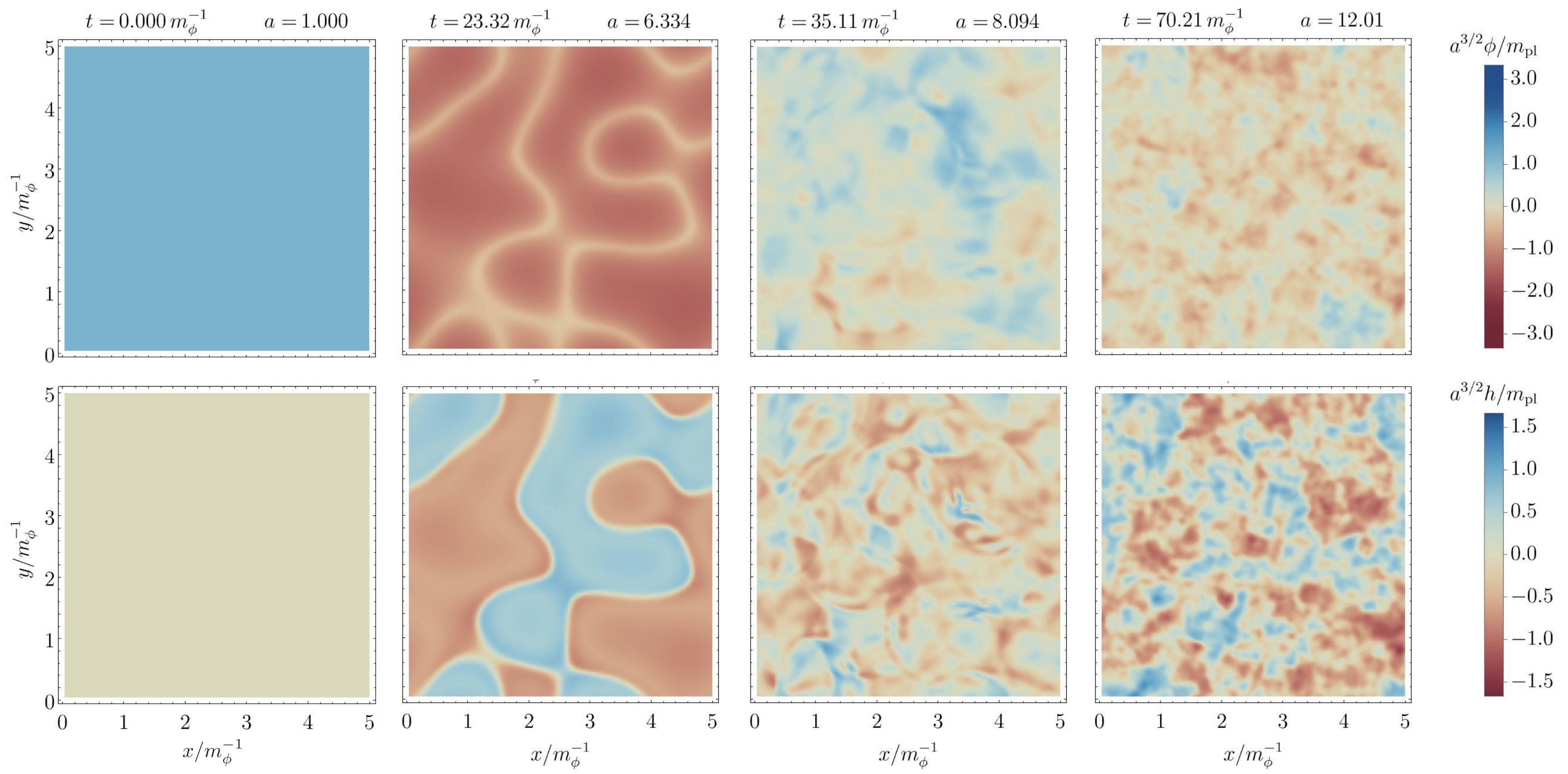}
\caption{Snapshots of the values of the modulus (first row) and Higgs (second row) fields on an arbitrary two-dimensional slice through the 3 dimensional simulation box at four different times (the spatial coordinates are co-moving). Around the time of backreaction, $t\approx23m_\phi^{-1}$ (second column), the Higgs field forms domains (`bubbles') with $h=\pm\sqrt{2|\phi|f}/q$. They disappear within $\Delta t\sim10m^{-1}$, due to collisions, as well as oscillations of the remnant $\phi$ condensate. The parameters we used are $\Delta=10^{-6},b = 0.9$, with $q=10^2$, $M=10^{-13}m_{\rm{pl}}$, $f=m_{\rm{pl}}$.}
%.} 
\label{fig:Snapshots}
\end{figure*}
%%~~~~~~~~~~~~~~~~~~~

%~~~~~~~~~~~~~~~~~~~
\begin{figure*}[t!] %  figure placement: here, top, bottom, or page
   \centering
   \includegraphics[width=7in]{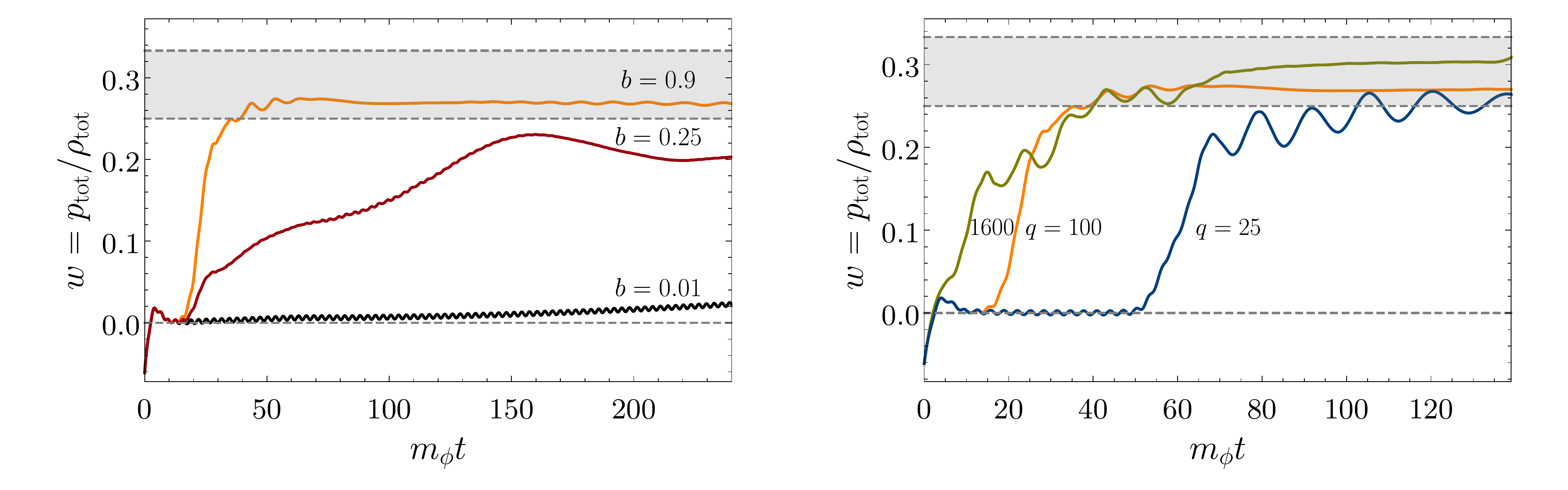} 
   \caption{Left Panel: Evolution of $w$ for the Higgs-modulus system for different values of the fragmentation efficiency parameter $b\equiv M^4/2\lambda f^2 m_\phi^2$ with tuning $\Delta = 10^{-6}$. For $b\sim \mathcal{O}[1]$, $1/4\lesssim w\lesssim 1/3$ is attained after fragmentation (orange curve). Smaller $b$ yields smaller late time $w$, with continued adiabatic evolution. In the untuned case ($\Delta\sim \mathcal{O}[1]$, not shown above) and $b\ne 1$, we get $w\approx 0$. Right Panel: For fixed $b=0.9$, varying $q=M^2/m_\phi^2$ affects when $1/4\lesssim w\lesssim 1/3$ is attained. For all curves, we have averaged energy densities and pressures spatially over the simulation box and temporally over fast oscillations.}
   \label{fig:EqOfState}
\end{figure*}
%~~~~~~~~~~~~~~~~~~~

%%%%%%%%%% --------- Nonlinear Dynamics --------- %%%%%%%%%%%%%%%
\section{Non-linear Dynamics}
%%%%%%%%%%%%%%%%%%%%%%%%%%
In a typical {\it untuned} scenario, when $m_\phi\gtrsim H$ in the early universe, the modulus field starts oscillating coherently along a valley of the potential, leading to an adiabatically evolving, early matter-domination epoch.

%\footnote{Gravitational clustering in cosmology occurs on timescales $\gtrsim H^{-1}$.}

In contrast, in a {\it tuned} universe, the modulus-Higgs system can undergo explosive, non-perturbative field dynamics leading to fragmentation of the fields on short time scales ($t\ll H^{-1}$), and yield a non-trivial equation of state for a number of $e$-folds of expansion following the fragmentation.

For $\Delta \ll 1$, the effective Higgs mass term oscillates between very large positive and negative values due to the oscillation of $\phi$. Such oscillations lead to non-adiabatic, out-of-equilibrium production of the Higgs particles (see Fig.~\ref{fig:b1r}). By considering tachyonic resonance \cite{Dufaux:2006ee}, for $f\sim \phi_{\rm in}\sim m_{\rm pl}$, the efficiency of such particle production is controlled by $q\equiv M^2/m_\phi^2$. In particular, $q\gg 1$ (as we assume) should lead to a broad range of physical momenta for the produced Higgs particles, see Fig.~\ref{fig:ps}.

Efficient transfer of energy from the modulus to the Higgs field is countered by the Higgs self-interaction $\lambda$. When particle production is sufficiently efficient, the Higgs field will be sufficiently populated in non-zero momentum modes to backreact on the modulus, making it spatially inhomogeneous (fragmented), as illustrated in Fig.~\ref{fig:Snapshots}. We will describe the process in more detail below.

%******************************************
\subsection{Does the modulus fragment?}
%******************************************
The Higgs field must be significantly populated to backreact and fragment the modulus. Large $q$ favors tachyonic resonance whereas large $\lambda$ limits the Higgs field occupation numbers. The parameter $b=M^4/{2\lambda f^2m_\phi^2}$ (introduced in eq.~\eqref{eq:bDef}) serves as a {\it fragmentation efficiency parameter}
since it incorporates both effects to determine whether the modulus fragments. At the level of the potential in eq.~\eqref{eq:toymodel}, $b$ controls the relative difference in the potential energy density between the ridge and valleys: $\Delta V= b\times (1/2) m_\phi^2(\phi-\phi_0)^2$. 

From detailed numerical simulations (\S\ref{supp:lattice}), we see no rapid fragmentation of the modulus field for $b\ll 1$; energetically, there is not much to be gained by falling into the valleys. For $b\sim \mathcal{O}[1]$, the modulus becomes completely fragmented, i.e.~the energy density in the zero mode of the modulus is comparable to that in high-momentum modes. We find that for the duration of our simulations after fragmentation, $\rho_h/\rho_\phi\sim 1$. That is, we are always left with significant energy density in the spatially inhomogeneous remnant modulus field (Fig.~\ref{fig:b1r}).  

Fig.~\ref{fig:ps} shows the power spectra of the two fields $P_F(k)\propto k^3|F(k)|^2$ ($F=h,\phi$) for understanding the distribution and time evolution of field perturbations at different scales. Note that the power spectra have been scaled by the  the amplitude of the oscillating modulus. Thus when the spectra are of order unity, the rms fluctuations in the fields are becoming comparable to the background modulus field, signaling fragmentation of the modulus.

Snapshots of the evolution of Higgs and modulus fields are shown in Fig.~\ref{fig:Snapshots}. The modulus first begins its oscillations from $\phi_{\rm in}=m_{\rm pl}$, then passes through $\phi=0$, causing the Higgs potential to develop minima. After a few oscillations, the fields start exploring these minima in a spatially inhomogeneous manner, leading to the formation of temporary domains. This is also the time when the backreaction on the oscillating modulus field becomes relevant. These domains quickly interact with each other and the still oscillating modulus field leading to complex spatio-temporal behaviour of the fields. The domains annihilate and the modulus field fragments spatially. The formation and dynamics of these domains turn out to be the dominant source of the gravitational wave signal we will discuss in \S\ref{subsec:stochasticGW} (see \S \ref{sec:gwlattice} for more details). We note that the existence of domain walls relies on there being a two dimensional field space. If the field space is higher dimensional, it is possible that higher dimensional transient defects like strings or textures will play a similar role.

The existence of transient $h$-domains (with accompanying domain walls) in this class of models is novel. 
Within a short period, $\Delta t\sim 10m_{\phi}^{-1}$, the domains disappear completely, and the fields enter a long turbulent stage. Perhaps, the shortness of the period in which the domains exist was the reason they were not noticed in \cite{Dufaux:2006ee}.

%******************************************
 \subsection{The Equation of State}
 %******************************************
The expansion history of an FRW universe is controlled by the equation-of-state parameter $w$:
\beq
w\equiv \langle p_{\rm tot}\rangle /\langle \rho_{\rm tot}\rangle\,,
\eeq
where $\langle \hdots \rangle$ indicates spatial averaging over $H^{-1}$ scales and temporal averaging over rapid oscillations in $p_{\rm tot}$. For fixed $b$, the detailed dynamics of the fields and time scale of fragmentation depend on the particular values of $q$ and $\lambda$. For example, for $b\sim\mathcal{O}[1]$, as $q$ increases, the modulus fragments earlier (Fig.~\ref{fig:EqOfState}, right panel). However, $w$ shows a simpler behavior as a function of $b$ in the tuned case when $\Delta \ll 1$:
\begin{itemize}
\item For $b\approx 1$, once fields fragment, we get $1/4\lesssim w\lesssim 1/3$ for the duration of our simulations ($\sim$ few $e$-folds). 
\item For $b\lesssim 1$, we find a non-trivial ($0<w<1/3$), adiabatically evolving $w$.
\item For $b\ll 1$, $w\rightarrow 0$. Again, we see some adiabatic evolution of $w$.
\end{itemize}
To sum up, along with $\Delta \ll 1$ (tuning), we also need $b\not\ll 1$ for significant nonlinear dynamics, fragmentation and a non-trivial ($w\ne 0$) equation of state (Fig.~\ref{fig:EqOfState}).

\subsection{Very Long-term Dynamics: Beyond Simulations}
We can only offer qualitative expectations for the long-term evolution of this highly nonlinear system. Even with complete fragmentation and an equation of state $w\sim 1/3$ seen in our simulations, significant energy density remains in the modulus field. We expect that after waiting long enough, without additional physics the universe will again become matter dominated.

Perturbative modulus decays occur on a timescale $\Gamma^{-1}\sim (m_{\rm pl}/m_{\rm \phi})^2 m_\phi^{-1} \gg m_\phi^{-1}$, much longer than the duration of the simulations ($t_{\rm sim}\sim {\rm few}\times 10^2 m_{\phi}^{-1}$). Energy could be drained more quickly from the modulus if the Higgs decays to other light species, freeing up phase space for further moduli conversion into the Higgs field. Plausibly, this might significantly reduce the energy density of the modulus compared to the decay products, though we have not simulated such dynamics. Nevertheless, it is difficult to see how matter domination can be avoided if even a small fraction of the initial energy density of the modulus survives in low momentum modes. In general we can allow a long-time averaged, constant $0<w_{\rm mod}<1/3$ to stand in for a range of possible behaviors (including the possibility of a nontrivial ($w\ne 0,1/3$) equation of state  maintained via nonlinear mode-mode couplings \cite{Dufaux:2006ee}).
\subsection{Without Fine Tuning}
So far we have focused on the fine-tuned case with $\Delta\ll 1$. For $\Delta \sim \mathcal{O}[1]$ and $b\not\ll 1$,
the modulus and Higgs fields can fall into the Higgs minima in a spatially inhomogenous manner. Nonlinear, spatially inhomogeneous field dynamics are thus possible even in theories that are not fine-tuned. However, we find that the fields end up in a state with an almost homogeneous modulus oscillating along one of the Higgs valleys around the global minimum. The initial Higgs production is typically not robust enough to backreact and break up the condensate. This behaviour quickly yields a standard matter-dominated phase with equation of state $w\approx 0$.  We have simulated $\Delta = 0.8$, $b = 0.5$ as well as $\Delta = 4 $, $b=0.9$ to see the behavior described above. We also found that for $\Delta =10^{-3}$ we see a transitionary behavior between the tuned and untuned case, with the equation of state evolving from radiation dominated towards matter domination during the duration of our simulations. These results confirm a general expectation that $\Delta$ controls the duration to matter domination, with smaller $\Delta$ leading to a longer duration.

Note that if $b=1$ (or sufficiently close to one), the existence of runaway directions changes the behavior qualitatively. In this case, we can end up with an equation of state $w\sim 1/3$ even when $\Delta\not \ll 1$. 

%~~~~~~~~~~~~~~~~~~~
\begin{figure}[t] %  figure placement: here, top, bottom, or page
   \centering
   \vspace{-15pt}
   \includegraphics[width=3.2in]{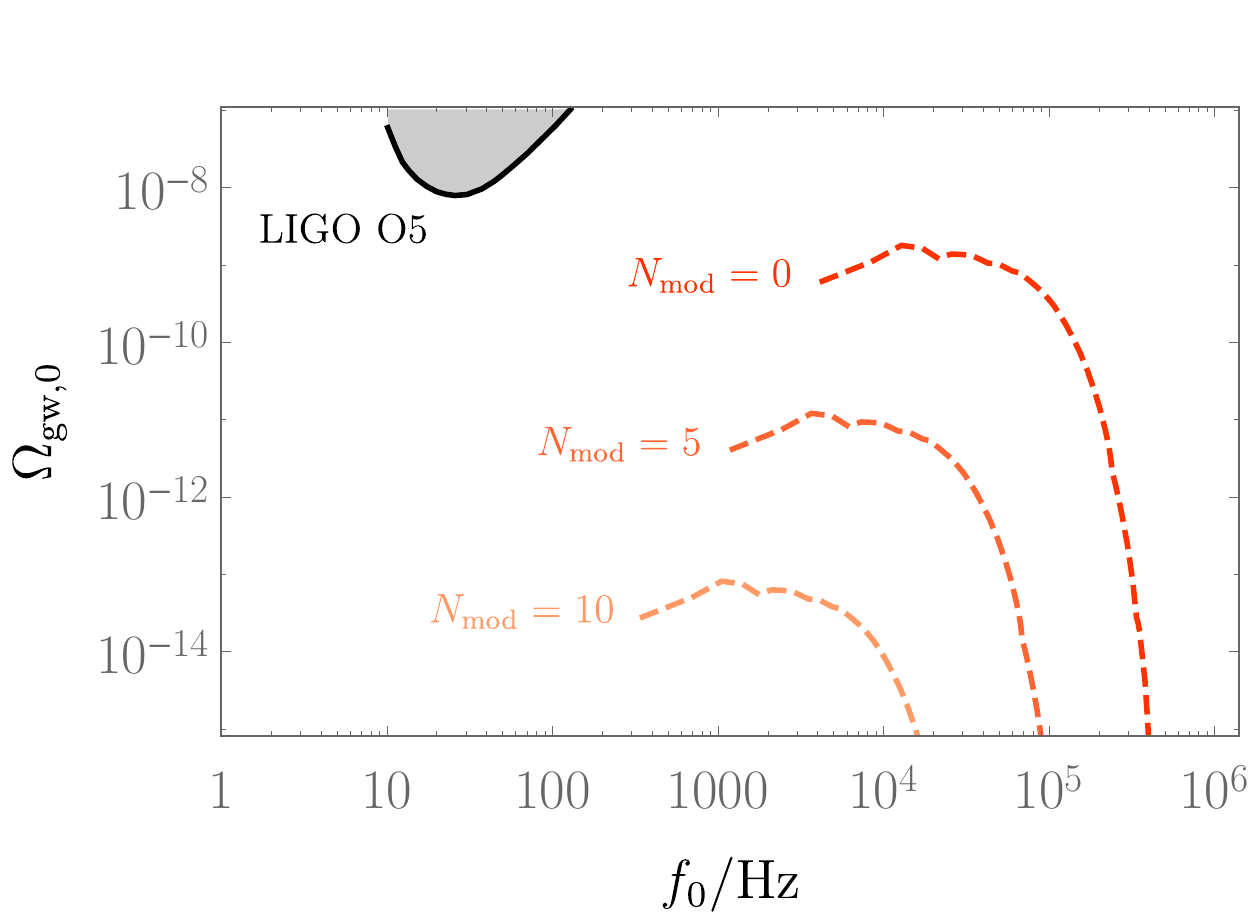} 
   \caption{Dashed orange curve with $N_{\rm mod}=0$: the gravitational waves (GWs) power spectrum today, generated by the non-linear dynamics at $t\approx70m_{\phi}^{-1}$ (assuming $\Delta =10^{-6},b=0.9,q=10^2, M/f=10^{-13})$. 
      The GWs on intermediate frequencies are generated by the slow propagation of power towards smaller comoving scales after backreaction; see Fig.~\ref{fig:ps}. Two paler dashed orange curves with $N_{\rm mod}>0$: rescaled versions of the top one, assuming $w_{\rm mod}=0$. Solid black curve: planned sensitivity of the fifth observational run of the aLIGO-AdVirgo collaboration \cite{TheLIGOScientific:2016wyq}.}
   \label{fig:GWs}
\end{figure}
%~~~~~~~~~~~~~~~~~~~
%%%%%%%%%%%%%%%%%% ------- Potential Signals and Consequences -------- %%%%%%%%
\section{Potential Signals and Consequences} 
%%%%%%%%%%%%%%%%%%%%%%%%%%%%%%%%%%%%%%%%%%%%%%%%%%%
%*********************************
\subsection{Stochastic Gravitational Waves} \label{subsec:stochasticGW}
%*********************************
For $b\not\ll1$, the fields in the modulus-Higgs system fragment rapidly (for $q\gg 1$), providing a source for the production of gravitational radiation \cite{Khlebnikov:1997di,Easther:2006gt,Dufaux:2007pt,GarciaBellido:2007dg}. The characteristic physical frequency of gravitational waves at the time of their generation is $f_{\rm g} \sim \beta^{-1} H_{\rm osc}$, with $\beta \sim  q^{-1/2}$ and $H_{\rm osc}\sim m_{\phi}$ the Hubble parameter when the modulus starts oscillating. Redshifting $f_{\rm g}$ to today, we obtain (see \S \ref{sec:gwlattice} for details)
\be
f_0\sim \frac{a_{\rm osc}}{a_0} \beta^{-1} H_{\rm osc} \sim \,{\rm kHz}\,\times\beta^{-1}\sqrt{\frac{m_\phi}{10 \, {\rm TeV}}}\,,
\ee
where we assume the universe can be approximated as radiation dominated shortly after $\phi$ begins oscillation. Note that for $\beta\ll 1$, these frequencies are beyond the reach of current interferometric detectors ($f_0\lesssim {\rm kHz}$), though not too far. Techniques for probing higher frequencies in the future have been discussed \cite{Akutsu:2008qv,Goryachev:2014yra, Arvanitaki:2012cn}.

The fraction of energy density in gravitational waves today (per logarithmic interval in frequency around $f_0$) can be estimated as \cite{Amin:2014eta}
\be
\Omega_{\rm gw,0} (f_0) \sim \Omega_{\rm r,0} \delta_\pi^2 \beta^2,
\ee
where $\Omega_{\rm r,0}$ is today's fraction of energy density stored in radiation and $\delta_\pi$ is the fraction of the energy density in anisotropic stresses when gravitational waves are produced. From the scalar field simulations (or estimates from linear instability calculations and energetic arguments), $\delta_\pi\sim 0.3$ and $\beta\sim q^{-1/2}$ which yield $\Omega_{\rm gw,0}\sim 10^{-8}$ for $q=10^2$. This result is consistent with our more detailed lattice simulations which calculate the gravitational wave spectrum using HLattice \cite{Huang:2011gf} (see Fig.~\ref{fig:GWs}). Note that detectable $\Omega_{\rm gw,0}(f_0\sim 10^2\rm Hz) \gtrsim 10^{-8}$ for aLIGO at design sensitivity \cite{TheLIGOScientific:2016dpb}. 

We can relax the assumption of a radiation-like equation of state immediately after fragmentation and generalize the above formulae. Assuming that (i) fragmentation and gravitational wave production happens quickly after modulus domination, (ii) the appropriately averaged equation of state $w=w_{\rm mod}$ for $N_{\rm mod}$ $e$-folds after fragmentation and before final radiation domination kicks in, the above formulae become
\begin{align}
f_0&\sim \,{\rm kHz}\times e^{-\frac{N_{\rm mod}}{4}(1-3w_{\rm mod})} \beta^{-1}\sqrt{\frac{m_\phi}{10 \, {\rm TeV}}}\,,\nonumber\\
\Omega_{\rm gw,0}(f_0)&\sim e^{-N_{\rm mod}(1-3w_{\rm mod})}\Omega_{{\rm r},0} \delta_\pi^2 \beta^2\,.
\end{align}
Note that a more observationally accessible, lower frequency signal using large values of $N_{\rm mod}(1-3w_{\rm mod})$ would lead to a significant suppression of $\Omega_{\rm gw,0}$, making detection challenging.  

A more coarse-grained constraint on total integrated gravitational wave energy density is provided by a measurement of the number of BSM light degrees of freedom present at the time of the CMB ($\Delta N_{\rm eff}$) through its impact on the cosmic microwave background \cite{Maggiore:1999vm}. Assuming gravitational waves are the only light degrees of freedom beyond those in the Standard Model, current constraints yield $\int d\ln f \,\,\Omega_{\rm gw,0}(f)\lesssim 10^{-6}$ \cite{Pagano:2015hma}, with an order of magnitude or more improvement expected from future missions \cite{Abazajian:2016yjj}. Our estimated $\int d\ln f \,\,\Omega_{\rm gw,0}(f)\sim 10^{-8}$ is within an order of magnitude of this future threshold,  and could potentially exceed it with either a wider scan of parameters, or inclusion of gauge fields \cite{Adshead:2018doq}. We note that this constraint does does not rely on the peak frequency of the gravitational waves, making larger $m_\phi$ acceptable.

%*********************************
\subsection{Constraints from/on Inflationary Observables}
%*********************************
Another possible consequence of the non-linear dynamics is to change the allowed $e$-folds during inflation. The $e$-folds between the time the current co-moving horizon scale exited the horizon during inflation and the end of inflation are related to the $e$-folds between the end of inflation and today in a given expansion history \cite{Liddle:2003as}. The expansion history also allows us to keep track of the evolution of the energy density. Then the $n_s$ and $r$ bounds from CMB measurements constrain an inflationary model together with its associated evolution afterwards. 

Assuming that during inflationary reheating, $w$ doesn't exceed $1/3$, we can obtain a conservative lower bound on $m_\phi$, 
\begin{align}
\frac{m_\phi^2}{m_{\rm pl}^2} \gtrsim & \exp\Bigg[\frac{-6 (1+w_{\rm mod})}{1-3w_{\rm mod}} \Bigg(57 - N_k + \ln \left(\frac{r \rho_k}{\rho_{\rm end}}\right)^{\frac{1}{4}}\Bigg)\Bigg]\nonumber
\end{align}
with $r$ the tensor-to-scalar ratio, $\rho_k$ ($\rho_{\rm end}$) the energy density when the mode exits the horizon (at the end of inflation), and $w_{\rm mod}$ the average $w$ between the time when the modulus starts oscillating and before it fully decays to radiation. For $0< w_{\rm mod} < 1/3$, the bound can be considerably weaker compared to when $w_{\rm mod} =0$. Details of the derivation and implications of this bound can be found in \S\ref{supp:inflation}.

%%%%%%%%%%%%%%%%%%%%%%%%%

\section{More Realistic Models} \label{sec:moremodel}

The simulation establishes that fragmentation requires $M^4 \sim \lambda m_\phi^2 f^2$. If we take $\lambda$ to be its Standard Model value, then $M \sim \sqrt{m_\phi f}$ and we cannot take both $M$ and $m_\phi$ of order the fundamental SUSY breaking scale. To make $b \sim 1$ compatible with the SM Higgs boson, one could take (for instance) $M \sim 10^{3}~{\rm TeV}$ and $m_\phi \sim 1~{\rm keV}$, or $M \sim 10^{11}~{\rm GeV}$ and $m_\phi \sim 100~{\rm TeV}$. However, the large mass hierarchy $M/m_\phi$ makes it very difficult to simulate the nonlinear dynamics. 

Closer to our simulations, we could take $m_\phi \lesssim M \sim 10^2~{\rm TeV}$ but $\lambda \sim 10^{-24}$. Then the $\lambda$ appearing in the simulation must differ from the observed $\lambda$ at the global minimum. This can happen in the SUSY two-Higgs doublet model, with its $D$-flat direction $|h_u| \approx |h_d|$ along which the effective quartic coupling can be tiny. If, as the modulus oscillates, the $D$-flat direction becomes tachyonic, we could achieve $b \sim \mathcal{O}[1]$. The modulus couplings must be arranged so that the point of marginal EWSB lies near the point at which the $D$-flat direction is accessible, which requires some additional fine-tuning. Loop corrections and higher dimension operators can produce effective quartic couplings $\lambda \sim m_{\rm soft}^2/\Lambda^2$, compatible with $b \sim 1$ if the cutoff $\Lambda \sim m_{\rm pl}$. Along the flat direction other SM particles become heavy, suppressing thermal effects. More details are in \S\ref{supp:modelbuilding}.

%%%%%%%%%%%%%%%%%%%%%%%%%%%%%%%%%%%
\section{Conclusions}
%%%%%%%%%%%%%%%%%%%%%%%%%%%%%%%%%%%

If the physical constants of the SM are determined by the VEVs of some scalar fields, in a tuned universe, even a small displacement of such a scalar field from its minimum can dramatically alter electroweak physics, leading to highly non-trivial dynamics in the early universe. We demonstrate this simple idea in a modulus-Higgs system.
We find that in the simplest model (eq.~\eqref{eq:toymodel}), for $b={M^4}/{(2\lambda f^2 m_\phi^2)}\sim \mathcal{O}[1]$, the fields fragment rapidly. This fragmentation leads to: (i) generation of gravitational waves; (ii) a non-trivial equation-of-state $1/4\lesssim w\lesssim 1/3$ for the duration of the simulations. The non-trivial equation of state can lead to a change in constraints on inflationary models, or alternatively, change constraints on the moduli mass. {\it Assuming} an equation of state $w\approx 1/3$ is maintained (for example, through the decay of the Higgs) up to eventual matter domination at $a_{\rm eq}$, we can expect a stochastic background of gravitational waves at high frequencies $f_0\gtrsim 10\,{\rm{kHz}}\times\sqrt{m_\phi/(10{\rm TeV})}$, with $\Omega_{\rm gw,0}\sim 10^{-9}-10^{-8}$.

This paper serves as the first step, and provides a template for, exploring the cosmological dynamics of electroweak fine tuning. We leave more realistic model building and numerical simulations (e.g.~the two Higgs doublet model with a flat direction) to future work.

\section*{Acknowledgments} We thank Patrick Draper and David Pinner for discussions in early stages of this project, Marcos Garcia for feedback on the manuscript, and Eiichiro Komatsu, Rouzbeh Allahverdi, Andrew Long, Vikram Rentala, and Mansi Dhuria for useful conversations. Part of the simulations were carried out on the COSMOS Shared Memory system at DAMTP, operated by U. of Cambridge on behalf of the STFC DiRAC HPC Facility. We thank D. Sijacki for her generosity regarding the use of her computational resources under the Cambridge COSMOS Consortium. For the rest of the simulations we used the facilities of the central computing center of the Max-Planck Society. We thank anonymous referees for helpful suggestions that have improved our presentation. This work was initiated at the Aspen Center for Physics, supported by National Science Foundation grant PHY-1607611. MA is supported by a DOE grant DE-SC0018216; JF by the DOE grant DE-SC0010010; MR by the NASA ATP Grant NNX16AI12G and by the DOE Grant DE-SC0013607.

\bibliography{draftRefs}

%%%%%%%%%%%%%%%%%%%%%%%%%%%%%%%
%%%%%%%%%%%%%%%%%%%%%%%%%%%%%%%
%%%%%%%%%%%%%%%%%%%%%%%%%%%%%%%

\onecolumngrid
\newpage
\newcommand\ptwiddle[1]{\mathord{\mathop{#1}\limits^{\scriptscriptstyle(\sim)}}}

\appendix 
\section{Field Dynamics and Lattice Simulations} \label{supp:lattice}
%*****************************************************
\subsection{Modulus--Higgs Potential}
%*****************************************************

\noindent We study a modulus $\phi$ and a Higgs field $h$ with the potential:
\begin{align}
V(\phi,h) &\equiv \frac{1}{2} m_\phi^2 (\phi - \phi_1)^2 + M^2 \frac{\phi - \phi_0}{f} h^\dagger h + \lambda (h^\dagger h)^2 + V_0.
\label{eq:toymodel2} 
\end{align}
As discussed in the main text, the field value $\phi_1$ denotes the minimum of the potential in the $\phi$ direction when $h=0$, whereas $\phi_0$ indicates the point of symmetry breaking. The global minimum of the potential is located at
\begin{align}
\phi_{\rm m} &= \frac{\phi_1 - b \phi_0}{1 - b} = \phi_0 - f \Delta, \\
|h_{\rm m}|^2 &= M^2 \frac{\phi_0 - \phi_1}{2 \lambda f(1-b)} = M^2 \frac{\Delta}{2 \lambda},
\end{align}
where  $\Delta\equiv (\phi_0-\phi_{\rm m})/f$ is the fine-tuning parameter defined in eq.~\eqref{eq:finetuning} and $b\equiv M^4/2\lambda m_{\phi}^2f^2$ is the fragmentation efficiency parameter defined in eq.~\eqref{eq:bDef}. Note that this minimum satisfies $\phi_{\rm m} < \phi_1$, and it is an electroweak symmetry breaking minimum (assuming $\phi_0 > \phi_1$). At this minimum the Higgs mass 
$m_h^2=2M^2\Delta$. For the case where the Higgs mass is light compared to its natural scale $M$, we need $\Delta\ll 1$. For $M\sim 10^2\,\rm{TeV}$, we take $\Delta\sim 10^{-6}$.
The additive constant in the potential in eq.~\eqref{eq:toymodel2},
\begin{equation}
V_0 \equiv \frac{b}{2(1-b)} m_\phi^2 (\phi_0 - \phi_1)^2 =\frac{1}{2}m_\phi^2 f^2\times b(1-b) \Delta^2,
\end{equation}
is chosen so that at the global minimum $V(\phi_{\rm m}, h_{\rm m}) = 0$.  

As noted in the text, for $\phi \leq \phi_1$ we can evaluate the potential energy difference between the ridge where $h = 0$ and the electroweak-breaking valley, i.e.~the minimum of the potential at fixed $\phi$. The result is
\begin{align}
\label{eq:DeltaVb}
\Delta V = b\times\frac{1}{2} m_\phi^2 (\phi - \phi_0)^2.
\end{align}
In particular, for small $b$ there is very little energy gained in rolling down from the ridge to the valley, and dynamical effects are suppressed. 

As seen from some of the above expressions, the case $b=1$ is to be handled with care. When $b \to 1$ at fixed $\phi_0 - \phi_1$, the global minimum runs away: $\phi_{\rm m} \to -\infty, |h_{\rm m}| \to \infty$. This behavior can be clarified by rewriting the potential as a sum of a positive definite term and a quartic Higgs potential:
\begin{align}
V(\phi,h) = \frac{1}{2} m_\phi^2 \left(\phi - \phi_1 + \frac{M^2}{m_\phi^2 f} h^\dagger h\right)^2 + \frac{M^2}{f} (\phi_1 - \phi_0) h^\dagger h + \lambda(1-b) (h^\dagger h)^2 + V_0.
\end{align}
This form of the potential makes it clear that if $\phi_1 = \phi_0$ and $b = 1$, the second and third terms vanish and there is a flat direction where $V = 0$ whenever $\phi \leq \phi_1, |h|^2 = m_\phi^2 f (\phi_1 - \phi)/{M^2}$. If $b = 1$ and $\phi_1 < \phi_0$, or if $b > 1$, the potential is not bounded below. Hence, we consider only the case $b < 1$ for which we have a well-defined global minimum. 

Without loss of generality, we can shift the field $\phi$ to set $\phi_1 = 0$. (Note that $\phi$ is uncharged, so $\phi = 0$ is not a special point in field space.) For simplicity, in our numerical simulations we treat $\phi$ and $h$ as two real scalar fields. The discussion above still applies after making the replacement $|h|^2 \to \frac{1}{2} h^2$. Hence, below we will work with the potential
\be
V(\phi, h) \equiv \frac{1}{2} m_\phi^2 \phi^2 + M^2 \frac{\phi - \phi_0}{2 f} h^2 + \frac{1}{4} \lambda h^4 + V_0. 
\label{eq:PotS1}
\ee

%*****************************************************
\subsection{Equations of Motion and Initial Conditions}
%*****************************************************
\noindent We work in a flat Friedmann-Robertson-Walker (FRW) universe with the metric
\beq
ds^2=dt^2-a^2(t)\delta_{ij}dx^idx^j\,.
\eeq
The dynamics of the modulus-Higgs system is determined by
\beq
\ddot{\phi}+3H\dot{\phi}-\frac{\nabla^2}{a^2}\phi+\partial_\phi V(\phi,h)=0\,,\qquad
\ddot{h}+3H\dot{h}-\frac{\nabla^2}{a^2}h+\partial_h V(\phi,h)=0\,,
\eeq
where the potential is given by \eqref{eq:PotS1}. The Hubble parameter is determined via the Friedmann equation with $H^2=(\dot{a}/a)^2=\langle \rho_{\rm tot}\rangle/3m_{\rm pl}^2$ where $\langle \rho_{\rm tot}\rangle$ is the spatially averaged, total energy density of the fields.

We note that for most of this section and the subsequent one on gravitational waves, we will provide results for the above toy model. Nevertheless, we will point out features that might be qualitatively different when considering the more realistic potential with a higher dimensional field space.

We assume that initially the modulus has a non-zero vacuum expectation value, $\phi_{\rm in}\sim f\sim m_{\rm pl}$ (where the Higgs has a positive mass), but that the Higgs does not, $h_{\rm in}=0$.\footnote{A more complete investigation of general initial conditions, especially in the negative Higgs mass regime, is left for future work.}  The initial Hubble rate is  (ignoring contributions from vacuum fluctuations)
\beq
H_{\rm in}\approx \frac{\sqrt{V_{\rm in}}}{\sqrt{3}m_{\rm pl}}= m_\phi\sqrt{\frac{1+b(1-b)\Delta^2}{6}}\,.
\eeq
Since the mass of the modulus is comparable to the Hubble rate, we expect the modulus to start oscillating right away. 

Along with the homogeneous fields, vacuum fluctuations ($\delta \phi$ and $\delta h$) are present in the fields. The mode functions for the quantum fluctuations satisfy linearized equations around a time-dependent classical background determined by $\phi(t)$  and $a(t)$. Such a linear description typically suffices to capture the initial evolution of $\delta \phi$ and $\delta h$. If there are growing (i.e., unstable) modes, the linear description eventually becomes inaccurate and the occupation number of these fields becomes quite high. Hence, it is plausible that the subsequent non-linear evolution system can be studied classically with lattice simulations.
%*****************************************************
\subsection{Linear instabilities in the Higgs}
%*****************************************************
The linearized equations of motion for $\delta\phi$ and $\delta h$ are
\begin{align}
\delta\ddot{\phi}+3H\delta\dot{\phi}-\frac{\nabla^2}{a^2}\delta\phi+m_{\phi}^2\delta\phi &=0\,, \\
\delta\ddot{h}+3H\delta\dot{h}-\frac{\nabla^2}{a^2}\delta h+\frac{M^2}{f}\left(\phi(t)-\phi_0\right)\delta h &=0\,,
\end{align}
implying that at the linear level the modulus fluctuations evolve as those of a scalar field with a constant mass, whereas the Higgs ones have a time-dependent mass which can lead to instabilities. We are primarily interested in $\Delta \ll 1$, with $\phi\sim f$, hence we ignore $\phi_0=(1-b)f \Delta $ compared to $\phi$ in the above equation for our instability analysis.

%~~~~~~~~~~~~~~~~~
\begin{figure*}[t] %  figure placement: here, top, bottom, or page
   \centering
 \includegraphics[width=6in]{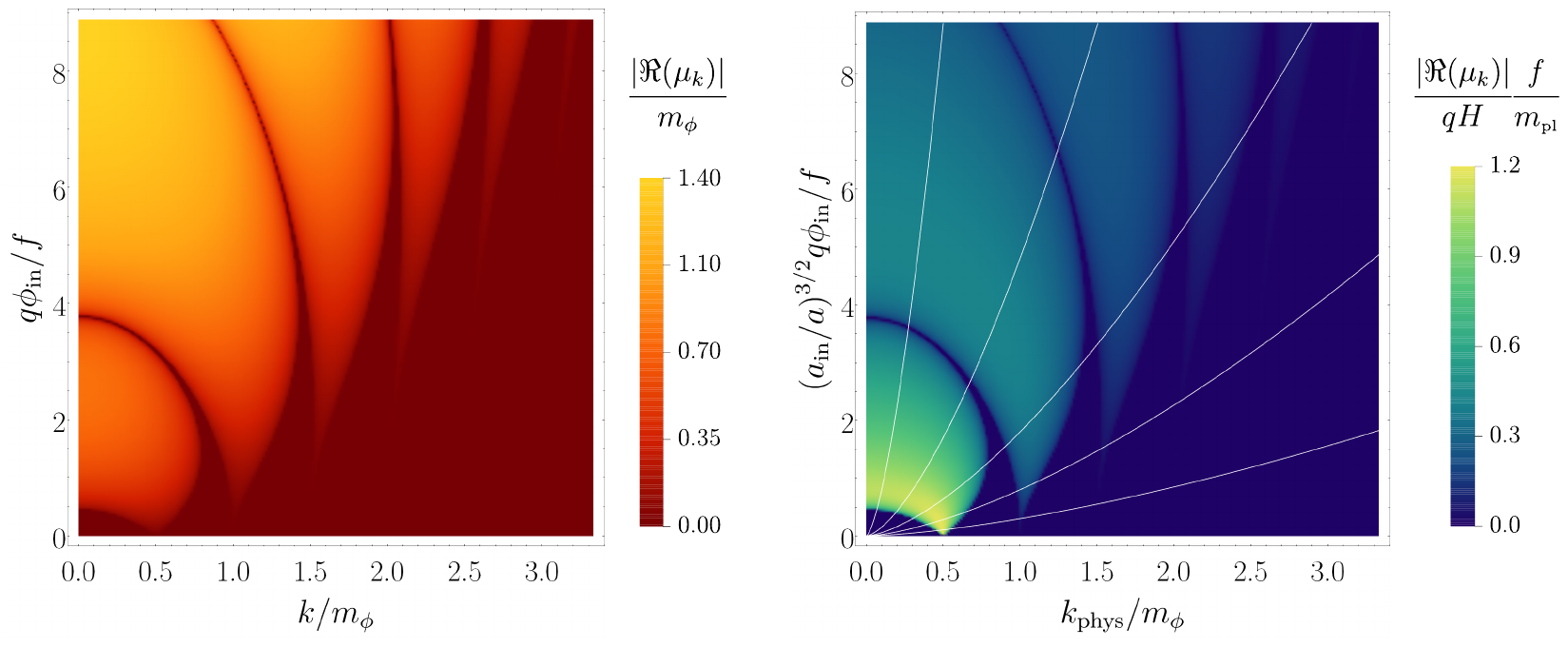} 
   \caption{The instability chart featuring the real part of the Floquet exponent normalized by the modulus mass (left) and the Hubble rate (right), characterizing the Higgs particle production rate. When $\phi_{\rm in}\sim f$, Higgs particle production is expected for $q>1$. In FRW space-time $k_{\rm phys}=k/ a(t)$, implying that a given co-moving mode flows towards the bottom left corner of the chart as the universe expands as indicated with the white lines in the second chart. Note that particle production is efficient if $|\Re(\mu_k)|/H\sim q m_{\rm pl}/f\gg1$.}
   \label{fig:FloqPhiChi2}   
\end{figure*}
%~~~~~~~~~~~~~~~~~

The Fourier modes of the canonically normalized Higgs, $\delta h_{\rm c}=a^{3/2}\delta h$, evolve according to
\beq
\label{eq:deltaHiggsEoM}
\delta\ddot{h}_{{\rm c}k}+\omega^2(k,t)\delta h_{{\rm c}k}=0\,,
\eeq
where
\beq
\omega^2(k,t)\approx \frac{k^2}{a^2}+\frac{M^2}{f}\phi(t)-(3H/2)^2-(3/2)\dot{H}\approx \frac{k^2}{a^2}+\frac{M^2}{f}\phi_{\rm in}\left(\frac{a_{\rm in}}{a(t)}\right)^{3/2}\cos(m_{\phi}t)\,.
\eeq
In the last line, we have used a standard approximation for a massive oscillating background scalar field, namely $a^{3/2}(t)\phi(t)\propto \cos(m_{\phi}t)$ and $3H^2\approx-2\dot{H}$. For small enough $k$
\beq
\label{eq:kcut}
\frac{k}{a(t)}=k_{\rm phys}<m_{\phi}\sqrt{q\frac{\phi_{\rm in}}{f}\left(\frac{a_{\rm in}}{a(t)}\right)^{3/2}}\,,
\eeq
This implies that $\omega^2(k,t)<0$ for nearly a half of the $\phi(t)$ oscillation. Such co-moving modes can then be unstable, and grow exponentially with time. In the context of {\it preheating} this amplification is known as {\it tachyonic resonance}.

To study parametric resonance in the Higgs from a periodic frequency change, one can resort to Floquet theory. If we ignore expansion, i.e., put $a(t)=\rm const.$ and $\phi(t)=\phi_{\rm in}\cos(m_{\phi}t)$, then Eq.~\eqref{eq:deltaHiggsEoM} is just the equation of motion of a simple harmonic oscillator with a periodically varying angular frequency. The Floquet theorem then tells us that its solution takes the form
\beq
\delta h_{{\rm c}k}(t)=e^{\mu_{k}t}\mathcal{P}_{k+}(t)+e^{-\mu_{k}t}\mathcal{P}_{k-}(t)\,,
\eeq
where $\mu_{k}$ is called the Floquet exponent and $\mathcal{P}_{k\pm}(t)$ are periodic functions of time. If $\Re(\mu_{k})\neq0$ one of the two terms increases exponentially with time. The numerically obtained exponent is given in the left panel in Fig.~\ref{fig:FloqPhiChi2} as a function of the model parameters. The broad instability bands are consistent with our naive expectations, Eq.~\eqref{eq:kcut}. To explain the additional features, such as narrow stability and instability bands, one has to consider the evolution of $\delta h_{{\rm c}k}(t)$ in greater detail, e.g., take into account the non-adiabatic change of $\omega^2(k,t)$ every time $\phi(t)=0$ for small enough $k$ and large initial amplitudes. 

However, these small features are irrelevant after the expansion of the universe is restored. In the right panel in Fig.~\ref{fig:FloqPhiChi2} we show that a given co-moving mode can flow across multiple broad instability bands. If $|\Re(\mu_k)|\gg H$, the mode amplitude can grow significantly within less than an {\it e}-fold of expansion. 

%*****************************************************
\subsection{Important Parameters for the Nonlinear Dynamics}
%*****************************************************
We have shown that the Higgs vacuum fluctuations can be linearly unstable and grow exponentially with time. As non-linear terms from the potential in Eq.~\eqref{eq:PotS1} become important, the exponential growth is expected to slow down. To estimate whether the energy in the amplified fluctuations is comparable to the background or not around the time non-linearities become significant we return to the {\it backreaction efficiency parameter}
\beq
b\equiv\frac{M^4}{2\lambda f^2m_{\phi}^2}=\frac{1}{4}\left(\cfrac{\cfrac{1}{2}\cfrac{M^2}{f}\phi h^2}{\cfrac{1}{2}m_{\phi}^2\phi^2}\right)\left(\cfrac{\cfrac{1}{2}\cfrac{M^2}{f}\phi h^2}{\cfrac{1}{4}\lambda h^4}\right)\leq1\,.
\eeq
As discussed above, $b < 1$ is required for $V\ge 0$. We have also ignored $\phi_0$ compared to $\phi$ for simplicity.

If $b\ll1$ and we assume that the energy in the amplified fluctuations is comparable to the background, i.e., $M^2\phi h^2/(2f)\sim m_{\phi}^2\phi^2/2$, then $M^2\phi h^2/(2f)\ll\lambda h^4/4$. The latter inequality implies that the quartic Higgs self-interaction has become important much earlier. Therefore, the Higgs instability is shut down before the amplified Higgs fluctuations have become energetic enough to backreact on the modulus background. We are left with a strongly self-coupled Higgs, interacting relatively weakly with the energetically dominant $\phi(t)$. The modulus is expected to remain homogeneous for a very long time.

If $b\lesssim1$ and we again assume that the energy in the amplified fluctuations is comparable to the background, i.e., $M^2\phi h^2/(2f)\sim m_{\phi}^2\phi^2/2$, then $M^2\phi h^2/(2f)\lesssim\lambda h^4/4$. The latter inequality implies that the quartic Higgs self-interaction becomes important around the time the amplified Higgs fluctuations have become energetic enough to backreact on the modulus background. The ensuing non-linear dynamics leads to the rapid fragmentation of $\phi(t)$.

As discussed earlier (see eq.~\eqref{eq:DeltaVb}), another related way of understanding the relevance of $b$ is as follows. The difference between the height of the ridge and the valleys in the potential is directly proportional to this same parameter $b$. As a result, $b\ll 1$ makes the potential energy gained by falling into the valleys negligible. Hence, a small $b$ suppresses significant non-linear dynamics from Higgs production and backreaction, consistent with the discussion above. In our simulations, we explore the dynamics of our system for $0.001\le b\lesssim 1$.

Another useful parameter that characterizes the nonlinear dynamics is $q\equiv M^2/m_\phi^2$ which controls the speed of energy transfer from the modulus to the Higgs (see right panel in Fig.~\ref{fig:EqOfState}). In our numerical investigations, we considered different $q$ in the range $25\le q\le 10^4$.

Note that for our simulations, we typically set $f\sim m_{\rm pl}$ and $M\sim 10^{-13}\,f$ and $q=M^2/m_\phi^2\sim 10^2$ and we increase/decrease these values by an an order of magnitude. With these sets of parameters, $b\sim1$ is achieved by choosing a very small $\lambda\ll 1$ (typically $\lambda\sim 10^{-24}$). However, qualitatively similar dynamics are expected even for large $\lambda$ (in particular for $\lambda \sim 0.1$ -- the SM value at the global minimum), as long as the other parameters are adjusted to still yield $b\sim 1$. To obtain $b\sim 1$ with $\lambda\sim 0.1$, we then need $M\sim \sqrt{m_\phi f}$. There are (at least) two possibilities to realize it in supersymmetric theories: low-scale SUSY breaking and very fine-tuned SUSY breaking, which we discuss further in section \S\ref{supp:modelbuilding}.

Our reason for not choosing these ``obvious" values ($\lambda\sim 0.1$) is that the time and length scale associated with tachyonic particle production ($\sim M^{-1}$) is extremely short compared to another natural time scale of the problem, $m_\phi^{-1}$ (the oscillation time-scale of the modulus). This disparity of scales creates a dynamical range problem for our simulations, and is beyond our ability to directly simulate given our computational resources.\footnote{For the largest  value of $\lambda$ used in our simulations ($\lambda = 10^{-18}$), our numerical time-step was $dt = 6.25\times 10^{-4}\,m_\phi^{-1}$ and spatial resolution was $dx=1.2\times 10^{-3}m_\phi^{-1}$. Our lattice had a size $N^3=512^3$ and we evolved our fields for $t=250\,m_\phi^{-1}$. For such a simulation, we required $\sim 10^4$ CPU hours. The time-step needed to resolve the tachyonic resonance scales as $dt\propto M^{-1}\propto \lambda^{-1/4}$. The same is true for the spatial resolution $dx$. Hence, increasing $\lambda$ to $0.1$ from $10^{-18}$ requires both reducing the time step and also increasing the spatial resolution by $\sim4$ orders of magnitude. With such small time steps, and high spatial resolution, simulating the field dynamics for $t\sim {\rm few}\times 100\, m_\phi^{-1}$ on a length scale of ${\rm few}\,m_\phi^{-1}$ will be beyond what is computationally feasible for us.

Changing $M/f$ or $q=M^2/m_\phi^2$ (by an order of magnitude each) while keeping $b$ fixed did not qualitatively change our results. The largest and smallest values of $\lambda$ we ran in our simulations while maintaining the same $b=0.9$ were $10^{-18}$ and $10^{-24}$ respectively. As expected, all these changes again did not affect our main claim: we get significant nonlinear dynamics, fragmentation and a non-trivial equation of state for $b\sim 1$ and $\Delta\ll 1$.}

%********************************************
\subsection{Lattice Simulations}
%********************************************
We use the parallelized version of LatticeEasy \cite{Felder:2000hq} to calculate the non-linear evolution of the fields and the self-consistent evolution of $a(t)$. The initial physical length of the edge of the simulation box is $L_{\rm{in}}=0.5 H_{\rm in}^{-1} -2.5 H_{\rm in}^{-1}$, whereas we set $a_{\rm{in}}=1$, with $a_{\rm end}\sim \mathcal{O}[\textrm{few {\em e}-folds}]$. Note that a slightly super-horizon box was needed sometimes to capture the tachyonic instability in $h$. The number of co-moving lattice points is $N=512^3$, and our time steps vary between  $dt = 0.00125 m_\phi^{-1}$ to  $0.000625 m_\phi^{-1}$ depending on the parameters chosen. The violation of the energy conservation in the above simulations is always less than $\mathcal{O}[10^{-4}]$.

At the start of the simulations $\phi$ has a background value, set to $\phi_{\rm{in}}=m_{\rm pl}$. The initial background field velocity, $\dot{\phi}_{\rm{in}}$, is equal to $-3H_{\rm{in}}\phi_{\rm{in}}/2$, in accordance with LatticeEasy conventions. The initial Fourier modes of the fields and field velocities (excluding the zero modes of $\phi$ and $\dot{\phi}$) are drawn from Gaussian probability distributions with covariance matrices equal to the squared amplitudes of the corresponding vacuum fluctuations. Initially, the energy budget is dominated by the homogeneous $\phi$, i.e., almost no energy is stored in the gradients. The values of $\phi_{\rm{in}}$ and $\dot{\phi}_{\rm{in}}$ imply that $w_{\rm{in}}\approx-1/4$ which is equivalent to starting the simulation soon after the end of slow-roll inflation if $\phi$ was the inflaton. \\

\begin{figure*}[t] %  figure placement: here, top, bottom, or page
   \centering
   \includegraphics[width=6.7in]{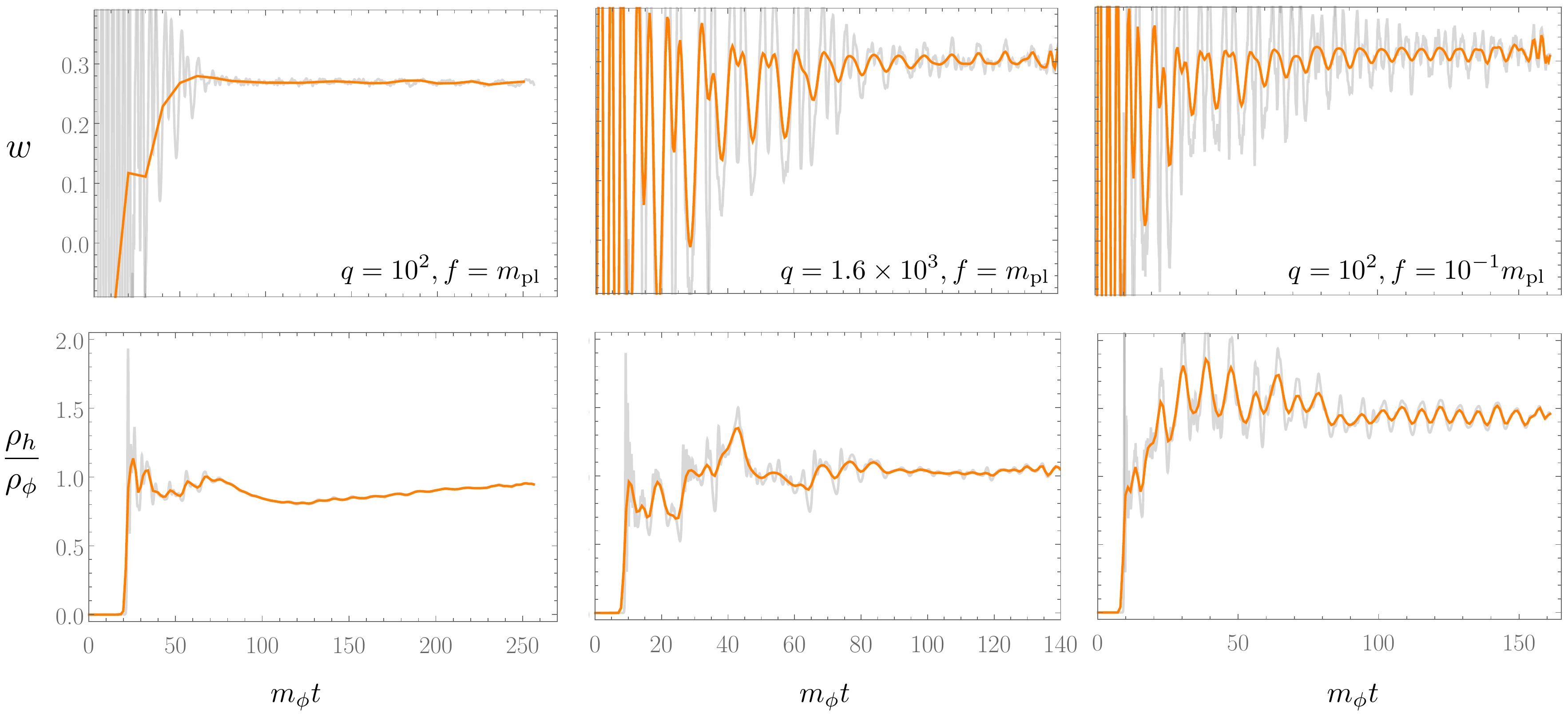} 
   \caption{The evolution of the equation of state, $w$, and the ratio of the mean Higgs and modulus densities, $\rho_{h}/\rho_{\phi}$. After backreaction, for $qm_{\rm pl}/f>10^2$, there is a short-lived oscillatory phase. Despite this curious behaviour $w$ settles to a constant value around $0.3$. We have chosen parameters such that $b=0.9$, $\Delta =10^{-6}$ in all cases. The grey and orange curves are obtained by averaging over space, with additional averaging over fast oscillations for the orange curves.}
   \label{fig:EoSSUPMAT}   
\end{figure*}

\noindent{\it Simulation Outputs}: 

Snapshots of the evolution of Higgs and modulus fields are shown in Fig.~\ref{fig:Snapshots}, along with the discussion of the dynamics in the main text. We do not repeat this discussion here. Along with the fields, we keep track of the spatially averaged energy density
\beq
\rho=\rho_{\phi}+\rho_{h}+\rho_{\rm{int}}+V_0\,,
\eeq
where
\beq
\rho_{\phi}=\frac{1}{2}\dot{\phi}^2+\frac{1}{2}\left(\frac{\nabla\phi}{a}\right)^2+\frac{1}{2}m_{\phi}^2\phi^2\,, \quad \rho_{h}=\frac{1}{2}\dot{h}^2+\frac{1}{2}\left(\frac{\nabla h}{a}\right)^2+\frac{1}{4}\lambda h^4\,,\quad \rho_{\rm{int}}=\frac{1}{2}\frac{M^2}{f}(\phi-\phi_0) h^2\,,
\eeq
as well as the pressure
\beq
p=\frac{1}{2}\dot{\phi}^2+\frac{1}{2}\dot{h}^2-\frac{1}{6}\left(\frac{\nabla\phi}{a}\right)^2-\frac{1}{6}\left(\frac{\nabla h}{a}\right)^2-\frac{1}{2}m_{\phi}^2\phi^2-\frac{1}{2}\frac{M^2}{f}(\phi-\phi_0) h^2-\frac{1}{4}\lambda h^4-V_0\,.
\eeq
The equation of state is defined as $w\equiv \langle p \rangle /\langle \rho\rangle$ where the angular brackets include a spatial average and when there are rapid oscillations, a temporal average as well. In Figs.~\ref{fig:EoSSUPMAT} and \ref{fig:EqOfState}, we show the results for the evolution of the energy densities and the equation of state for a range of parameters. Note that for the results in Figs.~\ref{fig:EoSSUPMAT}, we have chosen parameters so that the fragmentation efficiency $b = 0.9$, but allowed other parameters to vary. For the cases considered, the equation of state after fragmentation always settles near $1/4\lesssim w\lesssim 1/3$, and the amount of energy density in the modulus and Higgs fields are comparable.

%~~~~~~~~~~~~~~~~~
%~~~~~~~~~~~~~~~~~

%*****************************************************
\section{Gravitational Waves and Lattice Simulations}
\label{sec:gwlattice}
%*****************************************************
\subsection{Equations of Motion}
We calculate the gravitational waves generated by the nonlinear field dynamics using
\beq
\ddot{h}^{TT}_{ij}+3H \dot{h}^{TT}_{ij}-\frac{\nabla^2}{a^2}h^{TT}_{ij}=\frac{2}{m_{\rm pl}^2}\Pi_{ij}^{TT}
\eeq
where $h^{TT}_{ij}$ is the spatial, transverse, traceless part of the metric perturbations ($g_{\mu\nu}=g_{\mu\nu}^{\tiny \rm FRW}+h_{\mu\nu}$), and $\Pi^{TT}_{ij}$ is the transverse-traceless part of the energy momentum tensor of the fields which sources the gravitational waves. This is a ``passive calculation" where the (small) backreaction of the metric perturbations on the fields is ignored. 
\subsubsection{Characteristic Scales}
Let us consider a gravitational wave generated at $a=a_{\rm g}$ in the early universe with a co-moving wavenumber $k$. By taking into account red-shifting due to expansion and conservation of entropy after thermalization, the frequency today of this GW signal is
\beq
f_0=\frac{1}{2\pi}\frac{k}{a_{0}}=\frac{1}{2\pi}\left(\frac{k}{a_{\rm g}H_{\rm g}}\right)\sqrt{H_{\rm g} H_0}\left(\frac{a_{\rm g}}{a_{\rm th}}\right)^{(1-3{w}_{\rm mod})/4}\left(\frac{g_{\rm th}}{g_0}\right)^{-1/12}\Omega_{{\rm r},0}^{1/4}\,,
\eeq 
 where $H_{\rm g}$ is the Hubble parameter of the universe at the time of generation of the gravitational waves, $g_{\rm th}$ and $g_0$ are the effective number of relativistic degrees of freedom at the epoch of thermalization ($a_{\rm th}$) and today ($a_0$), $\Omega_{{\rm r},0}$ is the fractional energy density in relativistic species today and ${w}_{\rm mod}$ is the mean equation of state between generation and thermalization (after which we assume a standard thermal history). 

We can parametrize the characteristic wavenumber at which the gravitational waves are generated: 
\beq
\frac{k}{a_{\rm g}{H_{\rm g}}}\equiv\beta^{-1}\sim q^{1/2}\frac{m_{\rm pl}}{\sqrt{f\phi_{\rm g}}}\,,
\eeq
where the parameter $\beta$ has been estimated from an analysis of the linear instabilities in the field perturbations (see eq.~\eqref{eq:kcut}), with $\phi_{\rm g}$ being the amplitude of the modulus at the time of GW production.

The fraction of energy density in gravitational waves per logarithmic interval in wavenumber today is conventionally given as  $\Omega_{\rm gw,0}=\rho_{\rm c,0}^{-1}\left(d\ln\rho_{\rm gw,0}/{d\ln k}\right).$ Since GWs redshift as radiation, one can show that
\beq
\Omega_{\rm gw,0}=\Omega_{\rm gw}\times \left(\frac{a_{\rm g}}{a_{\rm th}}\right)^{1-3{w}_{\rm mod}}\left(\frac{g_{\rm th}}{g_0}\right)^{-1/3}\Omega_{{\rm r},0}\,,
\eeq
where $\Omega_{\rm gw}$ is the fractional energy density in gravitational waves at the time of generation. $\Omega_{\rm gw}$ can be estimated using the characteristic wavenumber above and assuming that a fraction $\delta_\pi$ of the energy density is involved in generating the gravitational waves (see for example \cite{Amin:2014eta}, with significant fragmentation, $\delta_\pi\lesssim 0.3$.):
\beq
\Omega_{\rm gw}=\frac{1}{\rho_{\rm g}}\frac{d\ln\rho_{\rm gw}}{d\ln k}\sim \beta^2\delta_{\pi}^2\,,
\eeq
where $\rho_{\rm g}$ is the total density at the time of generation of the gravitational waves. A more detailed discussion of such scalings (with slightly different parametrization) can be found \cite{Giblin:2014gra}.

For $g_{\rm th}/g_0=10^2$, $H_0=1.4\times10^{-33}\,\rm{eV}$, $\Omega_{{\rm r},0}=6.4\times10^{-5}$ \cite{Ade:2015lrj}, we can get an estimate of the characteristic frequency and amplitude of the gravitational energy density:
\begin{align}
f_0&\sim\beta^{-1}\sqrt{\frac{m_{\phi}}{10\,\rm{TeV}}}\sqrt{\frac{\phi_{\rm g}}{m_{\rm pl}}}\left(\frac{a_{\rm g}}{a_{\rm th}}\right)^{(1-3{w}_{\rm mod})/4}\times 1\,{\rm{kHz}}\,,\\
\Omega_{\rm gw,0}&\sim \beta^2\delta_\pi^2\left(\frac{a_{\rm g}}{a_{\rm th}}\right)^{(1-3{w}_{\rm mod})}\times 10^{-5}\,,
\end{align}
where $\beta^{-1}\sim q^{1/2} m_{\rm pl}/\sqrt{f\phi_{\rm g}}$. For the simulation parameters ($\Delta=10^{-6},q=10^2, b= 0.9,f=m_{\rm pl}$) for Figs.~\ref{fig:GWs} and \ref{fig:GWs1}, we get $\beta\sim0.1$.

\subsection{Lattice Simulations and Results}

To calculate the GWs we use HLattice \cite{Huang:2011gf}. We calculate them passively, i.e., we evolve the metric perturbations without accounting for their feedback on the fields and metric dynamics. We use the $6^{\rm th}$-order symplectic integrator for the self-consistent evolution of the fields and the scale factor, the HLATTICE2 spatial-discretization scheme and $k_{\rm eff}$ (not $k_{\rm std}$) for the TT projector. 

\begin{figure}[h]
  \begin{center}
%  \vspace{-10pt}
    \includegraphics[width=0.5\textwidth]{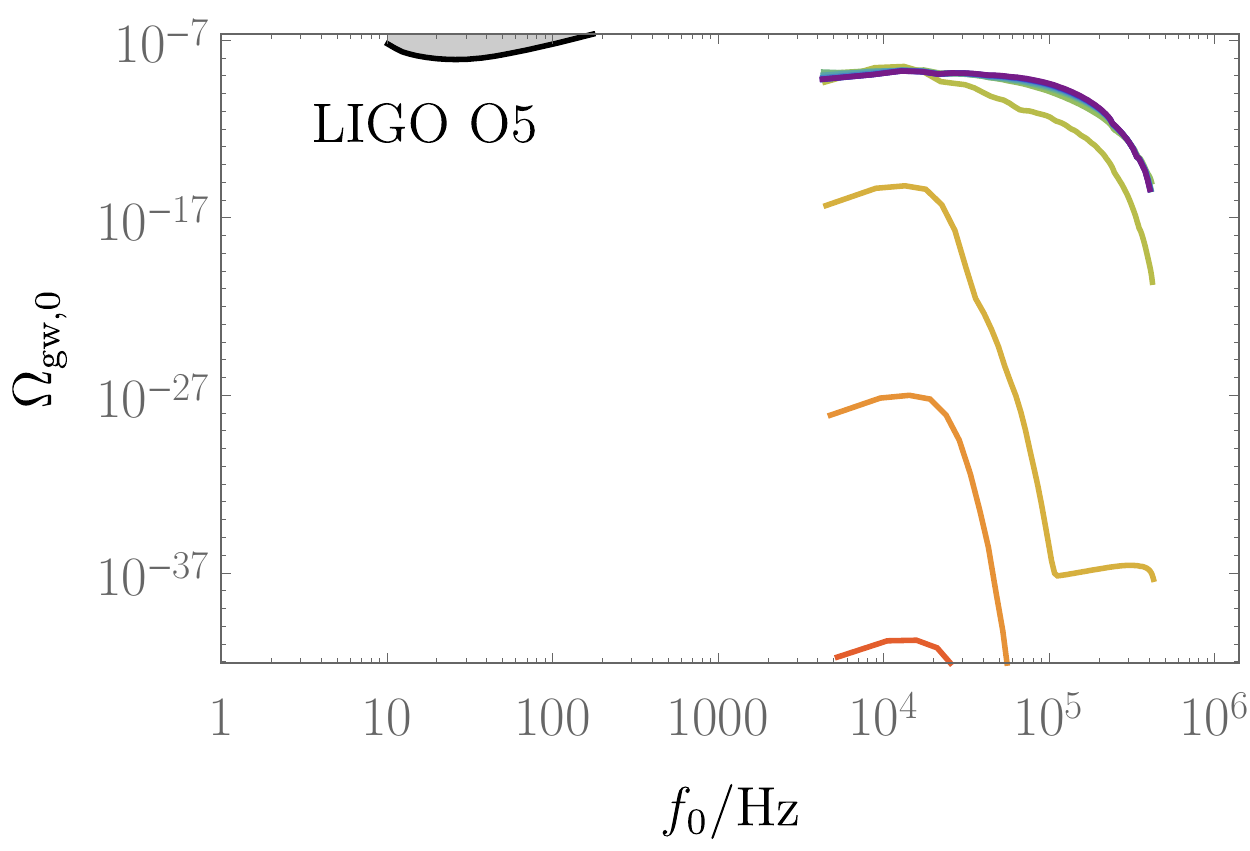}
  \end{center}
 %   \vspace{-10pt}
\caption{The growth in the amplitude of the GW power spectrum from the end of inflation to $t\approx70m_{\phi}^{-1}$(with $b=0.9, q=10^2, f=m_{\rm pl}$). The curves are output at time intervals $\Delta t=6m_{\phi}^{-1}$.}
%\vspace{-10pt}
\label{fig:GWs1}
\end{figure}

Figs.~\ref{fig:GWs} and \ref{fig:GWs1} are based on lattice simulations with $N=256^3$, $L_{\rm{in}}H_{\rm{in}}=2.0$ and $dt=L_{\rm in}/(16N^{1/3})\approx0.00120m_{\phi}$. The time step for the gravitational waves is $dt_{\rm GW}=4dt$. At the end of the simulation $a\approx12$, which corresponds to $t\approx70m_{\phi}^{-1}$ (this is also the time when the equation of state settles to a constant value, see orange curve in Fig.~\ref{fig:EqOfState}). 

The results of our simulations for gravitational waves are given in Fig.~\ref{fig:GWs1} (right). We show the time evolution of the gravitational wave spectra up to $t\approx70m_{\phi}^{-1}$.  The initial tachyonic instability in the Higgs generates GWs with well-defined cut-off, $f_0\lesssim q^{1/2}\sqrt{m_{\phi}m_{\rm pl}/(f\times10\,{\rm{TeV}})}\times \,{\rm{kHz}}\approx 10\,{\rm{kHz}}$, corresponding to the comoving modes $k<m_{\phi}q^{1/2}$. After backreaction, the spectrum settles down and GWs are slowly generated on intermediate frequencies, as power propagates towards smaller comoving scales, see Fig.~\ref{fig:ps}.
%~~~~~~~~~~~~~~~~~~~
%\begin{wrapfigure}{h}{0.5\textwidth}
%\end{wrapfigure}
%~~~~~~~~~~~~~~~~~~~

In Fig.~\ref{fig:GWs} in the main text, we scale the gravitational wave spectrum at $t\approx 70 m_\phi^{-1}$ assuming different subsequent expansion histories characterized by $(N_{\rm mod},w_{\rm mod})$. For the parameters $q=10^2,b =0.9,f=m_{\rm pl}$, we found $\delta_\pi\sim 0.3$ and $\beta\sim 0.1$, showing a consistency between our estimates in the previous sub-section and the results of the numerical simulations.

A more detailed understanding of the main source of gravitational wave production is obtained by specifically considering the domain walls formed in the Higgs-modulus system as seen in Fig.~\ref{fig:Snapshots}. The GW power emitted by a single `bubble' with quadrupole moment $Q$ and radius $R$ is (see \cite{GarciaBellido:2007af})
\beq
P_{\rm{gw,g}}\sim G\dddot{Q}^2\sim G\left(\frac{R^5\rho_{h\rm{,g}}}{R^3}\right)^2\,,
\eeq
where the subscript $\rm g$ denotes quantities at the time of generation of the GWs. We also have $P_{\rm{gw,g}}\sim \rho_{\rm{gw,g}}R^2$ from which follows that
\beq
\frac{\rho_{\rm{gw,g}}}{\rho_{h,{\rm{g}}}}\sim G\rho_{h,{\rm{g}}}R^2\,.
\eeq
At the time of domain formation $t_{\rm{g}}\sim22m_\phi^{-1}$, $\rho_{h,\rm{g}}\lesssim \rho_{\phi,\rm{g}}\lesssim\rho_{\rm{g}}$ (where $\rho_{\rm g}$ is the total energy density in the fields at the time of generation of the GWs). From the simulations $R\sim10m_{\phi}^{-1}$ (see second column in Fig.~\ref{fig:Snapshots}), implying
\beq
\Omega_{\rm{gw,g}}\sim\left(\frac{\rho_{h,\rm{g}}}{\rho_{\rm{g}}}\frac{\rho_{\phi,\rm{g}}}{\rho_{\rm{g}}}\right)\frac{\phi_{\rm{g}}^2}{m_{\rm{pl}}^2}\sim10^{-3}\,.
\eeq
In the above estimate, we take the factor in the brackets to be $\sim10^{-1}$ and $\phi_{\rm g}\sim 10^{-1}m_{\rm pl}$ consistent with simulations. This explains the strength of the signal $\Omega_{\rm{gw},0}\sim\Omega_{\rm{gw,g}}\times\Omega_{\rm r,0}\sim10^{-8}$. 

In our model with two real fields, the formation of the transient domain walls is important for the generation of GWs, giving an order of magnitude stronger signal than the one from the subsequent long turbulent stage. The time of formation of the domains and their length scale properly accounts for peak in the gravitational wave spectrum. In a more realistic theory, with a complex Higgs and moduli fields along with gauge fields, it is possible higher dimensional transient textures to play a qualitatively similar role. We leave this investigation to future work.

\section{Inflationary Constraints} \label{supp:inflation}
The key point is that the $e$-folds between the time the current comoving horizon scale exited the horizon during inflation and the end of inflation are related to the $e$-folds between the end of inflation and today in a given expansion history. The expansion history also allows us to keep track of the evolution of the energy density. Then the $n_s$ and $r$ bounds from CMB measurements constrain an inflationary model together with its associated evolution afterwards. This basic idea was proposed in Ref.~\cite{Liddle:2003as}. 

The cosmological history that we consider includes inflation, inflationary reheating characterized by a constant $w_{\rm re}$ in the equation of state,
radiation domination, an early matter domination phase starting when $H \approx m_\phi$ and the modulus begins to oscillate around its minimum, and radiation domination again after the perturbative decays of modulus. Differing from the discussions in Ref~\cite{Easther:2013nga, Dutta:2014tya, Cicoli:2016olq}, we include a possible non-trivial equation of state with a constant $w_{\rm mod} \neq 0$ originating from non-perturbative particle production after the modulus starts to oscillate and before the full conversion of the modulus energy into radiation.  The constant, $w_{\rm re}$, could be taken as an average from the end of inflation till radiation domination and satisfies 
\beq
\frac{\rho_{\rm rad}}{\rho_{\rm end}} = \left(\frac{a_{\rm end}}{a_{\rm rad}}\right)^{3(1+w_{\rm re})},
\eeq
where $a_{\rm end}, a_{\rm re}$ ($\rho_{\rm rad}, \rho_{\rm re}$) are the scale factors (energy densities) at the end of inflation and at the end of inflationary reheating respectively. 
Similarly, $w_{\rm mod}$ is the average from modulus oscillation till its full decay and satisfies
\beq
\frac{\rho_{\rm mod}}{\rho_{\rm dec}} = \left(\frac{a_{\rm dec}}{a_{\rm mod}}\right)^{3(1+w_{\rm mod})},
\eeq
where $a_{\rm mod}, a_{\rm dec}$ ($\rho_{\rm mod}, \rho_{\rm dec}$) are the scale factors (energy densities) when the modulus starts to oscillate and when full decays of the modulus happen (equivalently when radiation dominates again) respectively.

Our derivation closely follows Ref~\cite{Dutta:2014tya} and we will summarize the key steps below. 
The comoving Hubble scale $k=a_k H_k$ that exits the horizon during inflation could be written as
\beq
k= a_k H_k = \frac{a_k}{a_{\rm end}} \frac{a_{\rm end}}{a_{\rm re}} \frac{a_{\rm re}}{a_{\rm mod}}\frac{a_{\rm mod}}{a_{\rm dec}} a_{\rm dec}H_k,
\eeq
In terms of $e$-folds, $e^{N_k} = \frac{a_{\rm end}}{a_k}, e^{N_{\rm re}} = \frac{a_{\rm re}}{a_{\rm end}}, e^{N_{\rm RD}} =\frac{a_{\rm mod}}{a_{\rm re}}, e^{N_{\rm mod}}= \frac{a_{\rm dec}}{a_{\rm mod}}$, we have
\beq
\ln k = - N_{k} - N_{\rm re} - N_{\rm RD} - N_{\rm mod} + \ln a_{\rm dec} + \ln H_k. 
\eeq
Note that the $e$-folds between the modulus oscillation and full energy conversion into radiation is given by
\beq
N_{\rm mod} = \frac{1}{3(1+w_{\rm mod})} \ln \frac{\rho_{\rm mod}}{\rho_{\rm dec}}.
\eeq
In addition, $a_{\rm dec}$ could be rewritten in terms of the scale factor, $a_0$, today. Given the conserved comoving entropy, it can be achieved by relating the energy density at the end of modulus epoch, $\rho_{\rm dec}$ to the temperature today through 
\beq
\rho_{\rm dec} = \frac{\pi^2}{30} g_{\rm dec} T_{\rm dec}^4, \quad \frac{T_{\rm dec}}{T_0} = \left(\frac{g_{0;s}}{g_{{\rm dec};s}}\right)^{1/3} \frac{a_0}{a_{\rm dec}},
\eeq
where $g_{{\rm dec};s}$ and $g_{0;s}$ are the effective degrees of freedom for entropy. 
Furthermore, $N_{\rm RD}$ can be replaced by 
\begin{align}
\ln \rho_{\rm mod} &= \ln \frac{\rho_{\rm mod}}{\rho_{\rm re}} + \ln \frac{\rho_{\rm re}}{\rho_{\rm end}} + \ln {\rho_{\rm end}}  \\
&= -4 N_{\rm RD} - 3 (1+w_{\rm re}) N_{\rm re} + \ln \rho_{\rm end}
\end{align}
Combining all the equations above, we have 
\bea
\frac{1-3 w_{\rm mod}}{4} N_{\rm mod} &=&  - N_k - \frac{1-3 w_{\rm re}}{4} N_{\rm re} \nonumber \\
& +& \frac{1}{4} \ln \left( \frac{\pi^2}{30} g_{\rm dec}  \left(\frac{g_{0;s}}{g_{{\rm dec};s}}\right)^{4/3}\right) - \ln k + \ln H_k - \frac{1}{4} \ln \rho_{\rm end} + \ln (a_0 T_0)
\label{eq:Nmod1}
\eea
This equation relates the $e$-folds in the modulus epoch to the $e$-folds in the inflation epoch. For slow-roll inflation, 
\beq
H_k^2 = \frac{\pi^2}{2} m_{\rm pl}^2 r A_s = \frac{\rho_k}{3 m_{\rm pl}^2} \Rightarrow \ln H_k = \frac{1}{4} \ln \left(\frac{\pi^2 r A_s}{6} \right) +  \frac{1}{4} \ln \rho_k,
\label{eq:Hk}
\eeq
where $r$ is the tensor-to-scalar ratio, $A_s$ the amplitude of scalar perturbation and $\rho_k$ is the energy density when the mode exits the horizon. 
In addition, using 
\beq
\left(\frac{a_{\rm dec}}{a_{\rm mod}}\right)^{\frac{3}{2}(1+w_{\rm mod})} = 1 + \frac{3}{2} (1+w_{\rm mod}) H(t_{\rm mod}) (t_{\rm dec} - t_{\rm mod}),
\eeq
$N_{\rm mod}$ could be expressed in terms of the modulus mass,
\bea
N_{\rm mod} &\approx & \frac{2}{3(1+w_{\rm mod})} \ln \left(\frac{3}{2} (1+w_{\rm mod}) H(t_{\rm mod}) \tau_{\rm mod}\right), \nonumber \\
& = & \frac{2}{3(1+w_{\rm mod})} \ln \left(\frac{3}{2} (1+w_{\rm mod}) \frac{m_{\rm pl}^2}{c \times m_\phi^2}\right),
\label{eq:Nmod2}
\eea
where we approximated $t_{\rm dec}- t_{\rm mod}$ by the perturbative lifetime of the modulus $\tau_{\rm mod} = ({c m_\phi^3}/{m_{\rm pl}^2})^{\!-1}$ and $H(t_{\rm mod}) \approx m_\phi$. Putting Eq.~\eqref{eq:Nmod1}, \eqref{eq:Hk}, \eqref{eq:Nmod2} together, we have
\bea
&&\frac{1-3w_{\rm mod}}{6 (1+w_{\rm mod})} \ln \left(\frac{3}{2} (1+w_{\rm mod}) \frac{m_{\rm pl}^2}{c \times m_\phi^2}\right) = - N_k  - \frac{1-3 w_{\rm re}}{4} N_{\rm re} \nonumber \\
&+& \frac{1}{4} \ln \left( \frac{\pi^2}{30} g_{\rm dec}  \left(\frac{g_{0;s}}{g_{{\rm dec};s}}\right)^{4/3}\right)- \ln \left(\frac{k}{a_0T_0}\right)+ \frac{1}{4} \ln \left(\frac{\pi^2 r A_s}{6} \right) +  \frac{1}{4} \ln \left(\frac{\rho_k}{\rho_{\rm end}}\right) \\
&=&- N_k  - \frac{1-3 w_{\rm re}}{4} N_{\rm re} + 57+\frac{1}{4}\ln r + \frac{1}{4}\ln \left(\frac{\rho_k}{\rho_{\rm end}}\right),
\eea
where we use $\ln \left(10^{10} A_s \right) = 3.062$ (central value of Planck TT+lowP+lensing) at $k= 0.05$ Mpc$^{-1}$~\cite{Ade:2015lrj}, $T_0 = 2.725$ K, $g_{0;s} = 3.91$ and $g_{{\rm dec};s} = g_{\rm dec} = 10.76$. 
Thus we obtain a lower bound on $m_\phi$, 
\beq
m_\phi^2 \gtrsim \frac{3 (1+w_{\rm mod})}{2c} m_{\rm pl}^2 \exp\left(-\frac{6 (1+w_{\rm mod})}{1-3w_{\rm mod}}\left(- N_k - \frac{1-3 w_{\rm re}}{4} N_{\rm re}  + 57+\frac{1}{4}\ln r + \frac{1}{4}\ln \left(\frac{\rho_k}{\rho_{\rm end}}\right)\right)\right)
\label{eq:boundinflation}
\eeq
Note that generically we expect $0< w_{\rm re} < 1/3$ and $(1/4)(1-3 w_{\rm re}) N_{\rm re} >0$, which leads to a conservative bound on $m_\phi$ independent of the details of the inflation reheating stage
\beq
m_\phi^2 \gtrsim \frac{3 (1+w_{\rm mod})}{2c} m_{\rm pl}^2 \exp\left(-\frac{6 (1+w_{\rm mod})}{1-3w_{\rm mod}}\left(- N_k   + 57+\frac{1}{4}\ln r + \frac{1}{4}\ln \left(\frac{\rho_k}{\rho_{\rm end}}\right)\right)\right).
\label{eq:boundinflationcon}
\eeq
%~~~~~~~~~~~~~~~~~~~
\begin{figure}[t!] %  figure placement: here, top, bottom, or page
   \centering
   \includegraphics[width=3.2in]{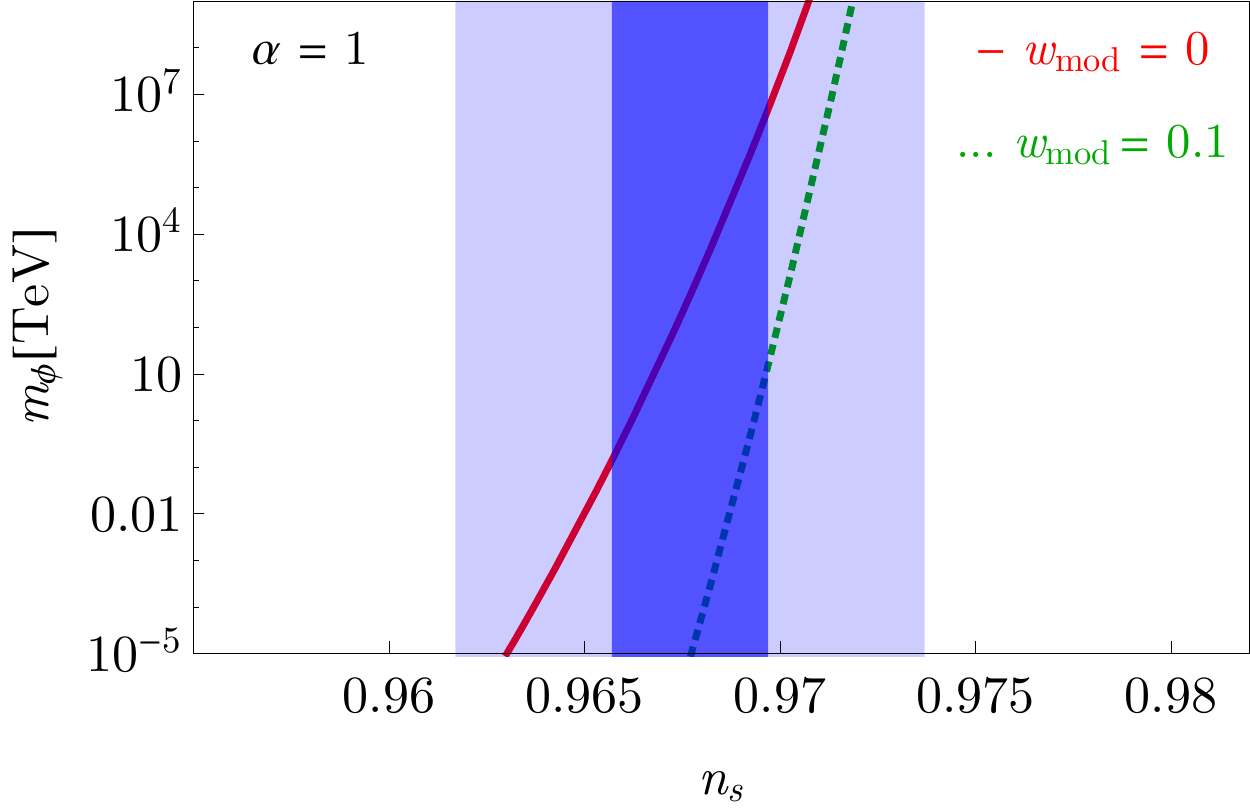} \quad    \includegraphics[width=3.2in]{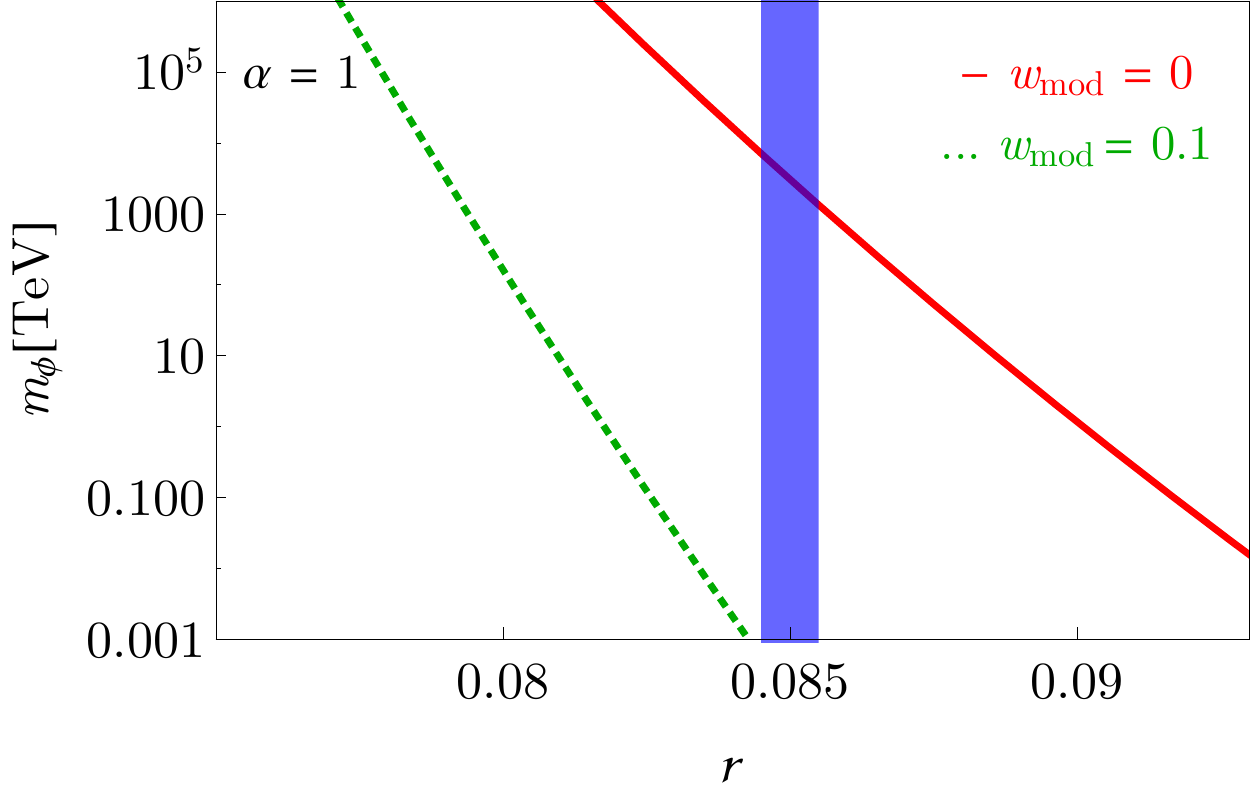} 
   \caption{The lower bound on $m_\phi$ as a function of $n_s$ (left) and $r$ (right) with the inflation model in Eq.~\ref{eq:inflation} and $\alpha = 1$. The red solid and green dotted lines correspond to $w_{\rm mod} = 0$ and 0.1 respectively. In the left panel, the light blue shaded region corresponds to the current 1$\sigma$ bounds on $n_s$ from Planck TT+lowP+lensing. The narrower darker blue shaded region corresponds to the $1\sigma$ bounds of a future CMB experiment of $n_s$ with sensitivity $\pm 2 \times 10^{-3}$~\cite{Abazajian:2016yjj}, assuming the same central value as Planck. In the right panel, the blue shaded region corresponds to the $1\sigma$ bounds of a future CMB experiment of $r$ with sensitivity $\pm 5\times 10^{-4}$~\cite{Abazajian:2016yjj}, assuming a measured central value of $r$ being 0.085.  }
   \label{fig:mphibound}
\end{figure}
%~~~~~~~~~~~~~~~~~~~

The presence of a non-zero $w_{\rm mod}$ could change the bound on $m_\phi$ dramatically compared to the case with $w_{\rm mod} = 0$. Since the logarithmic terms in the exponent in Eq.~\ref{eq:boundinflation}, \ref{eq:boundinflationcon} are usually tiny, a crude rule of thumb is that when $N_k < 57$, the bound could be significantly weakened with $w_{\rm mod} >0$ while when $N_k > 57.0$, the bound is more tightened with $w_{\rm mod} <0$. The details of the bounds depend on specific inflation models. Let's take a look at the model with a polynomial potential
\beq
V_{\rm inf}= \frac{1}{2}m^{4-\alpha}\phi_{\rm inf}^\alpha,
\label{eq:inflation}
\eeq
where $\phi_{\rm inf}$ is the inflaton and $\alpha > 0$. In this case, $N_k, r$ and ${\rho_k}/{\rho_{\rm end}}$ can be written in terms of the spectral index $n_s$ and the power $\alpha$: 
\begin{align}
N_k &= \frac{\alpha+2}{2(1-n_s)}, \quad r=\frac{8\alpha(1-n_s)}{\alpha+2} \\
\frac{\rho_k}{\rho_{\rm end}} &=\frac{2}{3} \left(\frac{2(\alpha+2)}{\alpha(1-n_s)}\right)^{\alpha/2}. 
\end{align}
In our evaluation below, we use $n_s = 0.9677 \pm 0.006$ (Planck TT+lowP+lensing)~\cite{Ade:2015lrj}. We also fix $c = {1}/{16\pi}$. For $\alpha = 1$, the lower bounds on $m_\phi$ as a function of $n_s$ or $r$ are illustrated in Fig.~\ref{fig:mphibound}. In this case, the central value of $n_s$ gives us 
$N_k \approx 46.4, r \approx 0.086, {\rho_k}/{\rho_{\rm end}} \approx 9$. This leads to a conservative lower mass bound of the modulus, $m_\phi > 477$ TeV when $w_{\rm mod} = 0$ and a much weaker bound when $w_{\rm mod}$ increases, e.g., $m_\phi > 8$ MeV when $w_{\rm mod} = 0.1$. Yet the potential strong mass bound on the modulus for $w_{\rm mod}=0$ may not be solid given the current precision of $n_s$. 
If we allow for $n_s$ to vary in the $1\sigma$ range, for instance, when $n_s$ takes the value at the lower $1\sigma$ bound, $n_s= 0.962$, $N_k \approx 39.2, r \approx 0.10, {\rho_k}/{\rho_{\rm end}} \approx 8.3$. When $w_{\rm mod} =0$, $m_\phi > 0.14$ MeV, which is negligible. In the future, if the precision of $n_s$ could be improved by a factor of 2 to 3 with the CMB-S4 measurements~\cite{Abazajian:2016yjj}, we will have a better assessment of the compatibility of the modulus scenario and different classes of inflation models. 

A more optimistic scenario is that in the near future, we will detect primordial gravitational waves and measure $r$. The precision of CMB-S4 measurement of $r$ is projected to be significantly improved to $5 \times 10^{-4}$. Assuming a measured $r = 0.085$ and CMB-S4's sensitivity, we could obtain a solid lower bound on $m_\phi$: $m_\phi > 1000$ TeV, when $w=0$ as shown in the right panel of Fig.~\ref{fig:mphibound}. When $w$ is increased to 0.1, the bound is considerably relaxed to be well below the cosmological moduli bound.

Additional cosmological constraints on this scenario could arise from isocurvature considerations~\cite{Iliesiu:2013rqa}. Alternatively, the field $\phi$ could itself be the inflaton, yielding additional constraints from the power spectrum of perturbations~\cite{Ade:2015lrj}.

%%%%%%%%%%%%%%%%%%%%%%%%%%%%%%%%%%%%%%%%%%%%%%%%%%%%%%%

\section{Aspects of the model}
\label{supp:modelbuilding}

\subsection{Approaches to $b \approx 1$}\label{supp:howtob1}

We have argued that the modulus fragments for a parameter choice
\begin{equation}
b \sim 1 \quad \Rightarrow \quad \lambda f^2 m_\phi^2 \sim M^4.
\end{equation}
As we will review below in \S\ref{supp:modulicouplings}, a standard scenario with moduli-mediated supersymmetry breaking will have both scalar mass parameters at the SUSY-breaking scale, $m_\phi \sim M \sim m_{3/2}$, and the modulus couplings suppressed by $f \sim m_{\rm pl}$. In that case, achieving $b \sim 1$ requires a tiny quartic coupling $\lambda \sim (m_{3/2}/m_{\rm pl})^2$. At first glance, this seems in conflict with the Standard Model Higgs quartic $\lambda \sim 0.1$. However, there are least three variations on this scenario that we can consider:
\begin{itemize}
\item Low-scale SUSY breaking: $M \sim 10^2~{\rm to}~10^3~{\rm TeV}$, $\lambda \sim 0.1$, $f \sim m_{\rm pl}$, $m_\phi \sim 10~{\rm eV}~{\rm to}~1~{\rm keV}$. Here the modulus is light because the fundamental scale of SUSY-breaking is low, but the Higgs mass scale is heavier due to stronger interactions with the SUSY-breaking sector.
\item Very fine-tuned SUSY breaking: $m_\phi \sim 10^2~{\rm to}~10^3~{\rm TeV}$, $\lambda \sim 0.1$, $f \sim m_{\rm pl}$, $M \sim 10^{11}~{\rm to}~10^{12}~{\rm GeV}$. Here we keep the modulus somewhat heavier than the TeV scale, but imagine that the natural scale for the Higgs VEV is orders of magnitude larger, closer to the intermediate scale. The physics is the same as the first case, except that the fundamental scale of SUSY breaking is larger and the weak scale is more fine-tuned (i.e.~$\Delta$ is much larger).
\item Proximity to a $D$-flat direction: $m_\phi \lesssim M \sim 10^2~{\rm to}~10^3~{\rm TeV}$, $f \sim m_{\rm pl}$, $\lambda \sim 10^{-24}$. In this case, a tiny effective quartic coupling is achieved along a $D$-flat direction. The Standard Model Higgs VEV does not lie along this direction, so the theory must be arranged so that our vacuum lies {\em near} the $D$-flat direction.
\end{itemize}

All of these three cases have interesting aspects, but none of them are completely trivial from the model-building viewpoint. In most of the remainder of this section we will focus on the last case, with a small quartic coupling along a $D$-flat direction. Our focus on this case is partly motivated by the fact that our simulations are all performed at very small $\lambda$, because the case $M \gg m_\phi$ is much more computationally expensive. Furthermore, because the Higgs field acquires very large values along a $D$-flat direction, most Standard Model particles will become very heavy and it may be a better approximation to neglect thermal effects in this case. Still, we think that all three of these scenarios are deserving of further exploration in the future.

\subsection{Origin of moduli couplings}\label{supp:modulicouplings}

In this section we will explain the origin of the $M^2 (\phi/f) h^\dagger h$ ansatz for the modulus coupling to the Higgs, and some variations that can arise. We first start by supposing that the modulus is a chiral superfield $\bm{X} \supset X + F_X \theta^2 $, with a supersymmetry breaking VEV
\begin{align}
\langle \bm{X} \rangle = X_0 + F_{X,0} \theta^2, \quad {\rm where} \quad X_0 \sim m_{\rm pl}, ~~F_{X,0} \sim m_{3/2} m_{\rm pl}.
\end{align}
Generic chiral superfields will obtain soft SUSY-breaking mass terms through couplings to $\bm{X}$,
\begin{align}
\int d^4 \theta \frac{\xi_{XZ}}{m_{\rm pl}^2} \bm{X}^\dagger \bm{X} \bm{Z}^\dagger \bm{Z} \supset \xi_{XZ} \frac{|F_X|^2}{m_{\rm pl}^2} Z^\dagger Z, \label{eq:XXZZ}
\end{align}
i.e.~$Z$ has a soft mass $ \sim m_{3/2}^2$. If $\bm{X}$ deviates from its vacuum expectation value, then in general this mass term will also fluctuate. For example, we might suppose that $\bm{X}$ has a superpotential
\be
W = \int d^2 \theta \left( \frac{1}{2} m_X \bm{X}^2 + \frac{1}{3!} g_X \frac{m_X}{m_{\rm pl}} \bm{X}^3 + \frac{1}{4!} \lambda_X \frac{m_X}{m_{\rm pl}^2}  \bm{X}^4 + \ldots\right),
\ee
where $g_X, \lambda_X \sim {\cal O}(1)$ and factors of $m_X/m_{\rm pl}^{k-2}$ have been extracted to ensure that $m_X$ acts as an overall spurion for shift-symmetry breaking. That is to say, it ensures that if $X \sim m_{\rm pl}$ all terms in the potential are of comparable size. Now, if $\bm{X}$ has a canonical K\"ahler potential $\int d^4 \theta \bm{X}^\dagger \bm{X}$, then we can solve for the $\theta^2$ component $F_X$ as:
\be
F_X^\dagger = \left(1 - \frac{\xi_{XZ}}{m_{\rm pl}^2} Z^\dagger Z + \ldots\right) \left(m_X X + \frac{1}{2} g_X m_X \frac{X^2}{m_{\rm pl}} + \frac{1}{3!} \lambda_X m_X \frac{X^3}{m_{\rm pl}^2} + \ldots\right).
\ee
From this we see that requiring that $X$ is the dominant source of SUSY breaking leads to $m_{3/2} \sim m_X$. This then parametrically guarantees that
\be
F_X \sim m_{3/2} m_{\rm pl} g(X)
\ee
where $g(X)$ is an order-one function of $X/m_{\rm pl}$. In particular, the term \eqref{eq:XXZZ} contains a trilinear coupling:
\be
\frac{2 \xi_{XZ} {\rm Re}(F_{X,0} m_X)}{m_{\rm pl}^2} {\rm Re}(X) Z^\dagger Z.
\ee
The prefactor here parametrically has size $m_{3/2}^2/m_{\rm pl}$. This is the analogue of our toy model, with $Z$ playing the role of the Higgs boson, ${\rm Re}(X)$ playing the role of the modulus $\phi$, and a prefactor of order $M^2/f$ with $f \sim m_{\rm pl}$ and $M \sim m_{3/2}$. In other words, a typical Planckian field displacement of $X$ from its minimum will lead to an order-1 variation in the soft mass of $Z$. 

We can also read off from this discussion that the $|F_X|^2$ term in the Lagrangian contains pieces that behave like
\be
\frac{\xi_{XZ}^2 |m_X|^2}{m_{\rm pl}^4} |Z|^4 |X|^2  \left(1 + {\cal O}(X/m_{\rm pl}) + \ldots\right).
\ee
In other words, we expect that moduli will inevitably generate quartic couplings of our fields with parametric size
\be
\lambda_Z \sim \frac{m_{3/2}^2}{m_{\rm pl}^2}.
\ee
Such $F$-term quartic couplings can also originate, as mentioned in the main text, from additional K\"ahler potential terms like $\int d^4 \theta \frac{\bm{X}^\dagger \bm{X}}{\Lambda^4} (\bm{Z}^\dagger \bm{Z})^2$. They will exist even, for instance, along $D$-flat directions of fields with gauge charges, as discussed in more detail below. The value of the quartic will be sensitive to the modulus value, but the parametric size will not.

In the context of the MSSM, moduli can affect Higgs soft masses by replacing $\bm{Z^\dagger Z}$ with  $\bm{h_{u,d}^\dagger h_{u,d}}$, or they can affect holomorphic ($b_\mu$-term) masses by coupling to $\bm{h_u h_d}$. If the modulus primarily affects the $b_\mu$-term rather than the soft masses, the dynamics can be rather different from our toy model, as a tachyonic direction exists both for large positive $b_\mu$ and for large negative $b_\mu$, possibly disappearing in an intermediate region as the modulus oscillates. It would be interesting to simulate this scenario in future work.

Many theories of moduli have special points in field space where the metric is singular and a tower of particles becomes light, e.g.~in string theory where many moduli fields $T$ have K\"ahler potentials of the form $a \log (\bm{T} + \bm{T^\dagger})$. Our field $\phi$ should be thought of as expanding around a value of $T \gg 1$, far from the singularity in moduli space at $T = 0$. The noncanonical K\"ahler term expanded around the minimum will give rise to terms like $\frac{1}{m_{\rm pl}^2} \phi^2 \partial_\mu \phi \partial^\mu \phi$, which may influence the dynamics. We assume that the field remains far from the singularity at $T = 0$, so that it is valid to work in terms of the canonically normalized field $\phi$. Nonetheless, as mentioned in \S\ref{sec:simplemodel}, the omitted terms could have important dynamical effects. It would be interesting to include such terms in future simulations.

In general, working with moduli whose imaginary parts have associated shift symmetries, which appear via the combination $\bm{T} + \bm{T^\dagger}$, does not qualitatively change the discussion. In certain sequestered scenarios, couplings may take a different form. For example, in the context of the large-volume scenario, we expect that the SM matter fields are sequestered from the overall volume modulus and the leading modulus decay is from the coupling \cite{Cicoli:2012aq, Higaki:2012ar}
\be
\int d^4 \theta \frac{\bm{\widetilde{T}} + \bm{\widetilde{T}^\dagger}}{\sqrt{3} m_{\rm pl}} \bm{h_u} \bm{h_d} + {\rm h.c.} \supset -\frac{1}{\sqrt{3} m_{\rm pl}} (\Box T) h_u h_d + {\rm h.c.}
\ee
Here $\bm{\widetilde{T}}$ is a modified chiral superfield missing its $F$-component, which is related to the conformal compensator in a superspace formulation of the theory \cite{Reece:2015qbf}. In the presence of an oscillating solution $\Box T \sim m^2 T$, this generates similar physics to a $b_\mu$ term linearly proportional to the modulus. After the modulus fragments, it could lead to rather different dynamics due to the derivatives acting on the modulus. Again, it could be interesting to simulate such variations in the future.

\subsection{The potential along a D-flat direction} \label{supp:Dterm}

Supersymmetric theories with renormalizable superpotentials generically have a variety of flat directions \cite{Affleck:1983mk,Affleck:1984xz}. The flat directions of the renormalizable, supersymmetric MSSM, together with the leading non-renormalizable operators that lift them, have been catalogued in \cite{Gherghetta:1995dv}. The existence of these flat directions is well known to have potential effects on cosmology, most famously for baryogenesis \cite{Affleck:1984fy, Dine:1995kz}.

Recall that in the MSSM, the tree-level Higgs mass matrix for the neutral modes $h^0_{u,d}$ takes the form
\be
\begin{pmatrix} |\mu|^2 + m_{h_u}^2 & -b_\mu \\ -b_\mu & |\mu|^2 + m_{h_d}^2 \end{pmatrix},
\ee
so it will have a tachyonic eigenvalue if one of the soft terms $m_{h_{u,d}}^2$ is sufficiently negative or if $b_\mu$ is sufficiently large (with either sign). We expect that in a sufficiently general theory, all of these terms will depend on the value of the modulus, so it oscillations can produce tachyons of either type (soft mass-driven or $b_\mu$-driven). There is a tachyonic SUSY-breaking mass along the supersymmetric D-flat direction $|h_u| = |h_d|$ when 
\be
m_{h_u}^2 + m_{h_d}^2 + 2|\mu|^2 - 2 |b_\mu| < 0.
\ee
This condition could arise dynamically as the modulus oscillates in many models, for instance those in which the $b_\mu$-term is driven by the $\phi$ oscillation. The condition may be especially easy to realize in models with an approximate shift symmetry that ensures $\tan \beta = 1$ at tree level \cite{Brignole:1996xb, Burdman:2002se, Choi:2003kq}, though this is not a necessary precondition. One might expect this tachyonic direction to be lifted by loop corrections; for example, there is a potential along the D-flat direction from one loop diagrams with tops or stops,
\be
V_{\rm 1-loop} \approx \frac{3 y_t^4 }{16\pi^2} (h_u^\dagger h_u)^2 \left[\log\frac{m_{\tilde t}^2}{m_t^2} + \frac{X_t^2}{m_{\tilde t}^2} \left(1 - \frac{1}{12} \frac{X_t^2}{m_{\tilde t}^2}\right)\right].
\label{eq:V1loop}
\ee
However, it is important to note that the masses $m_{\tilde t}$ and $m_t$ in this formula themselves depend on the value of the Higgs field, e.g.~$m_{\tilde t}^2 \approx y_t^2 |h^0_u|^2 + {\tilde m}^2_{Q_3, {\bar u}_3}$. At large values of the Higgs, EWSB contributions to the stop and top masses dominate over SUSY-breaking contributions and $\log\frac{m_{\tilde t}^2}{m_t^2} \sim m_{\rm soft}^2/|h|^2 \ll 1$. Effectively, far out along the flat direction supersymmetry is approximately restored in the sector of particles with large interactions with the Higgs boson. We can simply integrate them out, and the Higgs will behave as an approximate modulus with large field range. Similar results were discussed in \cite{Dvali:1998ct} in a finite-temperature context, where the presence of exponentially large values of MSSM fields in the early universe was argued to solve the monopole problem. (For a related discussion of zero-temperature physics, see the ``inverted hierarchy'' \cite{Witten:1981kv}.)

As is familiar from the Affleck-Dine mechanism, what will actually prevent the Higgs fields from taking arbitrarily large values along the flat direction are higher dimension operators.\footnote{In some cases, radiative effects will cause the tachyonic eigenvalue along the D-flat direction to run positive at values of the Higgs field well below the cutoff. It is then important to compute a renormalization group-improved effective potential. JF and MR thank Prateek Agrawal for useful conversations on this point, which we hope to explore in more detail elsewhere.} We can obtain quartic couplings along the flat direction from K\"ahler operators, for instance
\be
\int d^4 \theta \frac{\bm{X^\dagger X}}{m_{\rm pl}^4} (\bm{h_u^\dagger h_u})^2 \rightarrow \frac{|F_{X,0}|^2}{m_{\rm pl}^4} (h_u^\dagger h_u)^2.
\ee
This gives an effective quartic
\be
\lambda \sim \frac{m_{3/2}^2}{m_{\rm pl}^2},
\ee
which is precisely what is needed to give a fragmentation efficiency $b \sim 1$, assuming $m_\phi, M \sim m_{3/2}$ and $f \sim m_{\rm pl}$.

At first glance it appears that superpotential terms can prevent such large field values. For example, a superpotential
\be
\int d^2 \theta \left(\mu \bm{h_u h_d} + \frac{1}{M_*} (\bm{h_u h_d})^2\right)
\ee
gives rise to quartic terms such as 
\be
\frac{\mu^\dagger}{M_*} (h_u^\dagger h_u)(h_u h_d) + {\rm h.c.},
\ee
which would stop the Higgs along the flat direction at values of order $(\mu M_*)^{1/2}$. If we take $M_* \sim m_{\rm pl}$, these are small field values and we would never achieve a sufficiently large fragmentation efficiency. However, any realization of the MSSM should contain a solution to the $\mu$ problem, explaining why the coefficient of $\int d^2\theta \, \bm{h_u h_d}$ is much smaller than the Planck scale. We expect that such a solution will generically imply that higher order superpotential terms like $\int d^2\theta \, (\bm{h_u h_d})^2$ also have parametrically small coefficients related to the same spurion $\mu/m_{\rm pl}$. Provided that $1/M_* \lesssim \mu/m_{\rm pl}^2$, we obtain a sufficiently small quartic.

Since this spurion argument is rather abstract, let us consider a more explicit example of the expected size of the Higgs quartic coupling in the context of a particular solution of the $\mu$ problem. The Giudice-Masiero mechanism \cite{Giudice:1988yz} invokes a K\"ahler term $\int d^4 \theta \left(\frac{c_\mu}{m_{\rm pl}} \bm{X}^\dagger \bm{h_u} \bm{h_d} + {\rm h.c.}\right)$ which, if the $F$-component of $\bm{X}$ obtains a VEV, becomes an effective superpotential $\mu$-term with size of order soft SUSY-breaking parameters. For this mechanism to work, it is necessary that the true $\mu$-term $\int d^2\theta\, \mu \bm{h_u h_d}$ be highly suppressed or altogether absent from the superpotential. Although one can invoke the supersymmetric nonrenormalization theorem to excuse this assumption as technically natural, a better approach is to invoke a symmetry explanation (approximate or exact, discrete or continuous). For example, concrete completions of the Giudice-Masiero mechanism invoking discrete, anomaly free $R$-symmetries exist \cite{Babu:2002tx, Lee:2011dya}. As a simple example, the $\mathbbm{Z}_4$ $R$-symmetry under which the superpotential has charge 2, the matter fields $\bm{q, \bar u, \bar d, \ell, \bar e}$ have charge 1 and the Higgs fields $\bm{h_{u,d}}$ have charge 0 suffices to forbid a $\mu$-term and enforce matter parity for proton stability. Notice that this symmetry forbids not only the $\mu$-term itself but also higher-dimension operators such as $\int d^2 \theta \frac{1}{\Lambda} (\bm{h_u h_d})^2$ that could affect the Higgs quartic coupling. In the context of this $\mathbbm{Z}_4$ symmetry, we will encounter terms like
\be
\int d^4 \theta \frac{c_{\mu,2}}{m_{\rm pl}^3} \bm{X}^\dagger (\bm{h_u h_d})^2 \to \int d^2 \theta \frac{c_{\mu,2}F_{X,0}^\dagger}{m_{\rm pl}^3} (\bm{h_u h_d})^2,
\ee
an effective superpotential quartic term with coefficient $\sim \frac{\mu}{m_{\rm pl}^2}$. In other words, if the role of Giudice-Masiero is to suppress the $\mu$ term relative to the Planck scale by a small spurion $\mu/m_{\rm pl}$, the discrete symmetry approach ensures that the quartic Higgs superpotential term is suppressed by the same small spurion.

\subsection{Proximity to a D-flat direction} \label{supp:proxDflat}

We have argued that the effective quartic coupling for the Higgs boson can be very small when the tachyonic direction of the potential is aligned with the $D$-flat direction. However, to fit low-energy Standard Model physics, we would like to have an effective quartic $\lambda \approx 0.1$ for the light Higgs mode at the global minimum. One can then ask if it is plausible that the oscillation of a modulus in the early universe is able to probe the $D$-flat direction for a long period of time. Achieving this requires an extra condition: not only do we need the global minimum of the potential to be near the point of marginal EWSB (the condition for our vacuum to be fine-tuned), we also need the global minimum to be near the point in field space at which the $D$-flat direction becomes tachyonic. The proximity of {\em three} special points in field space amounts to an extra fine-tuning, beyond the usual one. On the other hand, if our vacuum has $\tan \beta$ near 1, the amount of additional fine-tuning may be small.

As an example of how modulus couplings might probe the flat direction, consider a scenario where as a function of the modulus $\phi$ the three Higgs potential parameters $M_{H_d}^2 \equiv |\mu|^2 + m_{H_d}^2$, $M_{H_u}^2 \equiv |\mu|^2 + m_{H_u}^2$, and $b_\mu$ have the dependence
\begin{align}
M_{H_{u,d}}^2 \equiv (\alpha_{u,d} \phi/m_{\rm pl} + \beta_{u,d}) M_S^2, \quad b_\mu \equiv (\phi/m_{\rm pl}) M_S^2,
\end{align}
where $M_S$ is a measure of SUSY-breaking and $\alpha_{u,d}, \beta_{u,d}$ are dimensionless parameters. If $\alpha_u \alpha_d > 1$, then at $\phi \gg 0$, there is no tachyon and the symmetry is unbroken. On the other hand if $\alpha_{u,d} > 0$, then at $\phi \ll 0$, there is a tachyon along the $D$-flat direction and the Higgs can acquire large values. Thus, qualitatively the picture of unbroken electroweak symmetry on one side and badly broken electroweak symmetry on the other is similar to the toy model we have simulated. As argued in the preceding subsection, the effective quartic coupling can be very small at $\phi \ll 0$.

The theory deviates more from our toy model in the region in between unbroken electroweak symmetry and electroweak symmetry broken badly along the flat direction. The point of marginal electroweak breaking is when $\phi^2/m_{\rm pl}^2 = (\alpha_u \phi/m_{\rm pl} + \beta_u)(\alpha_d \phi/m_{\rm pl} + \beta_d)$, whereas the condition for a tachyon to point along the $D$-flat direction is $|\phi|/m_{\rm pl} > \frac{1}{2}\left[(\alpha_u + \alpha_d) \phi/m_{\rm pl} + (\beta_u + \beta_d)\right]$. If $\alpha_u \approx \alpha_d$ and $\beta_u \approx \beta_d$ then these are approximately the same condition, and tuning the point of marginal electroweak breaking to lie near the global minimum will ensure that the evolving modulus provides access to the flat direction at a nearby point in field space. However, in the Standard Model, where the up-type Higgs couples to the top quark much more strongly than the down-type Higgs couples to the bottom quark, radiative corrections would tend to spoil such a relationship. Hence, proximity to a $D$-flat direction is likely to require some additional fine-tuning beyond the tuning of the Higgs boson mass itself. A full study of loop corrections in such a scenario, and how much fine-tuning is required, is beyond the scope of this work.

\end{document}